\documentclass[twocolumn,pre]{revtex4-1}
\usepackage{epsfig}
\usepackage{amssymb,amsfonts,amsmath}
\usepackage[pdftex]{pstricks}

\include{epsf}

\newcommand{\beq}{\begin{equation}}
\newcommand{\eeq}{\end{equation}}
\newcommand{\bea}{\begin{eqnarray}}
\newcommand{\eea}{\end{eqnarray}}

\newcommand{\eqn}[1]{\begin{equation}#1\end{equation}}

\begin{document}

\title{Beyond position weight matrices: nucleotide correlations in transcription factor binding sites and their description}

\author{Marc Santolini}

\address{Laboratoire de Physique Statistique, CNRS,
    Universit\'e P. et M. Curie, Universit\'e D. Diderot,
    \'Ecole Normale Sup\'erieure,  Paris, France.}
 
\author{Thierry Mora} 

\address{Laboratoire de Physique Statistique, CNRS,
    Universit\'e P. et M. Curie, Universit\'e D. Diderot,
    \'Ecole Normale Sup\'erieure,  Paris, France.}

\author{Vincent Hakim}

\address{Laboratoire de Physique Statistique, CNRS,
    Universit\'e P. et M. Curie, Universit\'e D. Diderot,
    \'Ecole Normale Sup\'erieure,  Paris, France.}

\begin{abstract}
The identification of transcription factor binding sites (TFBSs) on genomic DNA is of crucial importance for understanding and predicting regulatory elements in gene networks.
TFBS motifs are commonly described by
Position Weight Matrices (PWMs), in which each DNA base pair independently contributes 
 to the transcription factor (TF) binding, despite mounting evidence of interdependence between base pairs positions.
The recent availability of 
genome-wide data on TF-bound DNA regions 
offers the possibility to revisit this question in detail for TF binding {\em in vivo}. Here, we use available 
fly and mouse ChIPseq data, and show that the independent model generally does not reproduce the observed statistics of
TFBS, generalizing previous observations. We further show that TFBS description and predictability can be systematically
improved by taking into account pairwise correlations in the TFBS via the principle of maximum entropy. The resulting pairwise interaction model is formally equivalent to the disordered Potts models of statistical mechanics and it generalizes previous approaches to interdependent positions. Its structure allows for 
co-variation of two or more base pairs, as well as secondary motifs. Although models consisting of mixtures of PWMs also have this last feature, we show that pairwise interaction models outperform them.
The significant pairwise interactions are found to be sparse and found dominantly between consecutive base pairs.
Finally, the use of a pairwise interaction model for the identification of TFBSs is shown to give significantly different predictions than a model based on independent positions.
\end{abstract}

\maketitle

\section*{Author Summary}

Transcription factors are proteins that bind on DNA to regulate several processes such as gene transcription or epigenetic modifications. Being able to predict the Transcription Factor Binding Sites (TFBSs) with accuracy on a genome-wide scale is one of the challenges of modern biology, as it allows for the bottom-up reconstruction of the gene regulatory networks. The description of the TFBSs has been to date mostly limited to a simple model, where the affinity of the protein for DNA, or binding energy, is the sum of independent contributions from uncorrelated amino-acids bound on base pairs. However, structural aspects are of prime importance in proteins and could imply appreciable correlations throughout the observed binding sequences. Using a statistical physics inspired description
and high-throughput ChIPseq data for a variety of Drosophilae and mammals TFs, we show that such correlations exist and that accounting for their contribution greatly improves the predictability of genomic TFBSs.  

\section*{Introduction}

Gene regulatory networks are at the basis of our understanding of a cell state and of the dynamics of its response to environmental cues. Central effectors of this regulation are Transcription Factors (TF) that bind on short DNA regulatory sequences and interact with the transcription apparatus or with histone-modifying proteins to alter target gene expressions \cite{spitz12nrg}. 
The determination of Transcription Factor Binding Sites (TFBSs) on a genome-wide scale 
is thus of
importance and is the focus of many current experiments \cite{encodesurvey12}. An important feature of TF in eukaryotes is that their binding specificity is moderate and that a given TF is found to bind a variety of different sequences {\em in vivo} \cite{wasserman}.
The collection of binding sequences for a TF-DNA is widely described by a Position Weight Matrix (PWM) which simply gives the probability that a particular
base pair stands at a given position in the TFBS. The PWM provides a full statistical description of the TFBS collection when there are no correlations between nucleotides
at different positions. Provided that the TF concentration is far from saturation, the PWM description applies exactly at
 thermodynamic equilibrium in the simple case where the different nucleotides in the TFBS
contribute independently to the TF-DNA interaction, such that
the total binding energy is the mere sum of the individual contributions \cite{Berg:1987p12416, Stormo:1998p12421}. 

Previous works have reported 
several cases of correlations between nucleotides at different positions in TFBSs \cite{Man:2001p12827,Benos:2002p12828,Bulyk:2002p12734,jolmataipale13cell}. A systematic {\em in vitro} study of 104 TFs using DNA microarrays 
revealed a rich picture of binding patterns \cite{Badis:2009p12808}, including the existence of multiple motifs, strong nucleotide position interdependence, and variable spacer motifs, where two small determining regions of the binding site are separated by a variable number of base pairs. Recently, the specificity of several hundred human and mouse DNA-binding domains was investigated using high-throughput SELEX. Correlations between nucleotides were found to be widespread among TFBSs and predominantly located between adjacent flanking bases in the TFBS \cite{jolmataipale13cell}.
The relevance of nucleotide correlations remains however debated \cite{zhao11natbiot}.

On the modeling side, probabilistic models have been proposed to describe these correlations, either by explicitly identifying mutually exclusive groups of co-varying nucleotide positions \cite{Benos:2002p12828,Zhou:2004p12739,Hu:2010p12741}, or by assuming a specific and tractable probabilistic structure such as Bayesian networks or Markov chains 
\cite{barash2003mod,Sharon:2008p12737, jolmataipale13cell}.
However, the extent of nucleotide correlations in TFBSs {\em in vivo} remains to be assessed, and a systematic and general framework that accounts for the the rich landscape of 
observed TF binding behaviours is yet to be applied in this context. 
The recent breakthrough in the experimental acquisition of precise, genome-wide TF-bound DNA regions with the ChIPseq technology offers the opportunity to address these two important issues.
Using a variety of ChIPseq experiments coming both from fly and mouse, we first show that the independent model generally does not reproduce well the observed TFBS statistics
for a majority of TF. This calls for a refinement of the PWM description that accounts for interdependence between nucleotide positions.

The general problem of devising interaction parameters from observed state frequencies has been recently studied in different contexts where large amounts of data have become available. These include describing the probability
of coinciding spikes \cite{Schneidman:2006p2295, Shlens:2006p1442} or activation sequences \cite{ikegaya,roxin08} in neural data, 
the statistics of protein sequences
\cite{Weigt:2009p3341,Mora:2010p5398}, and even the flight directions of birds in large flocks \cite{Bialek:2012p12539}. 
Maximum-entropy models accounting for pairwise correlations in the least constrained way have been found to provide significant improvement over independent models. 
The PWM description of TF binding is equivalent to the maximum entropy solely constrained by nucleotide frequencies at each position. Thus, we propose, in the present paper, to refine this model by further constraining pairwise correlations between nucleotide positions.
 This corresponds to including effective pairwise interactions between nucleotides in an equilibrium thermodynamic model of TF-DNA interaction, as already proposed \cite{zhaostormo12gen}.
When enough data are available, the TFBS statistics and predictability are found to be significantly improved in this refined model. 
We consider, for comparison, a model that describes the statistics of TFBSs as a statistical mixture of PWMs \cite{barash2003mod} and generalizes previous proposals
\cite{Cao:2010fk,Heinz:2010fk}. This alternative model can directly capture some higher-order correlations between nucleotides but is found
to be outperformed for all considered TF by the pairwise interaction model. 

We further show that the pairwise interaction model accounts for the different PWMs appearing in the mixture model by studying its energy landscape: each
basin of attraction of a 
metastable energy minimum in the pairwise interaction model is generally dominantly described by one PWM in the mixture model.
Significant pairwise interactions between nucleotides are sparse and found dominantly between consecutive nucleotides, in 
general qualitative agreement with {\em in vitro} binding results \cite{jolmataipale13cell}.
The proposed model with pairwise interactions only requires a modest computational effort. 
When enough data are available, it should thus generally prove worth using the refined description of TFBS that it affords.

\section*{Results}

\begin{figure*}[ht]
\begin{center}
\includegraphics[width=\linewidth]{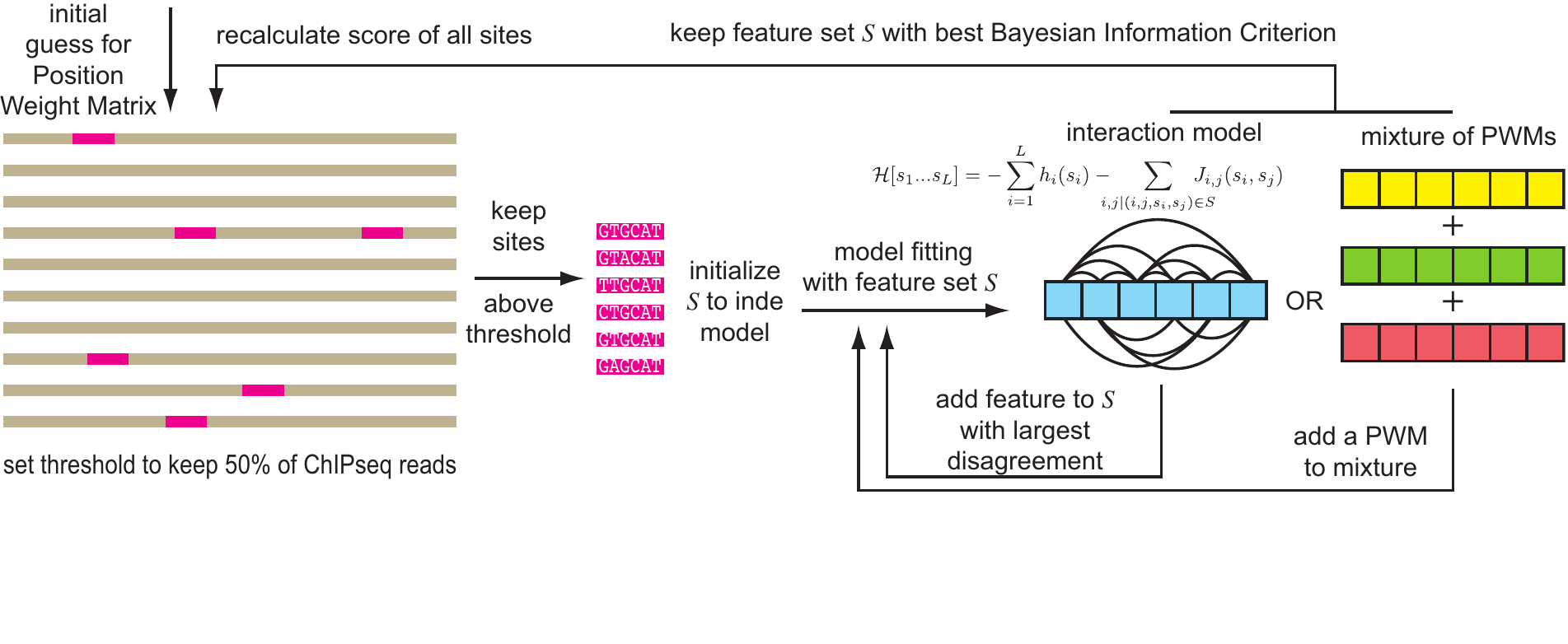}
\caption{
{\bf Workflow.} An initial Position Weight Matrix (PWM) is used to find a set of binding sites on ChIPseq data. Models are then learned using single-point frequencies (independent), two-point correlations (pairwise) or a mixture of independent models learned on sites clustered by K-Means (mixture) with increasing complexity, {\em i.e.} increasing number of features in the model. Finally the models with best Bayesian Information Criteria (BIC) are used to predict new sites until convergence to a stable set of sites. 
}
\end{center}
\label{fig:workflow}
\end{figure*}

\subsection*{The PWM model does not reproduce the TFBS statistics}
\begin{figure*}[ht]
\begin{center}
\includegraphics[width=.8\linewidth]{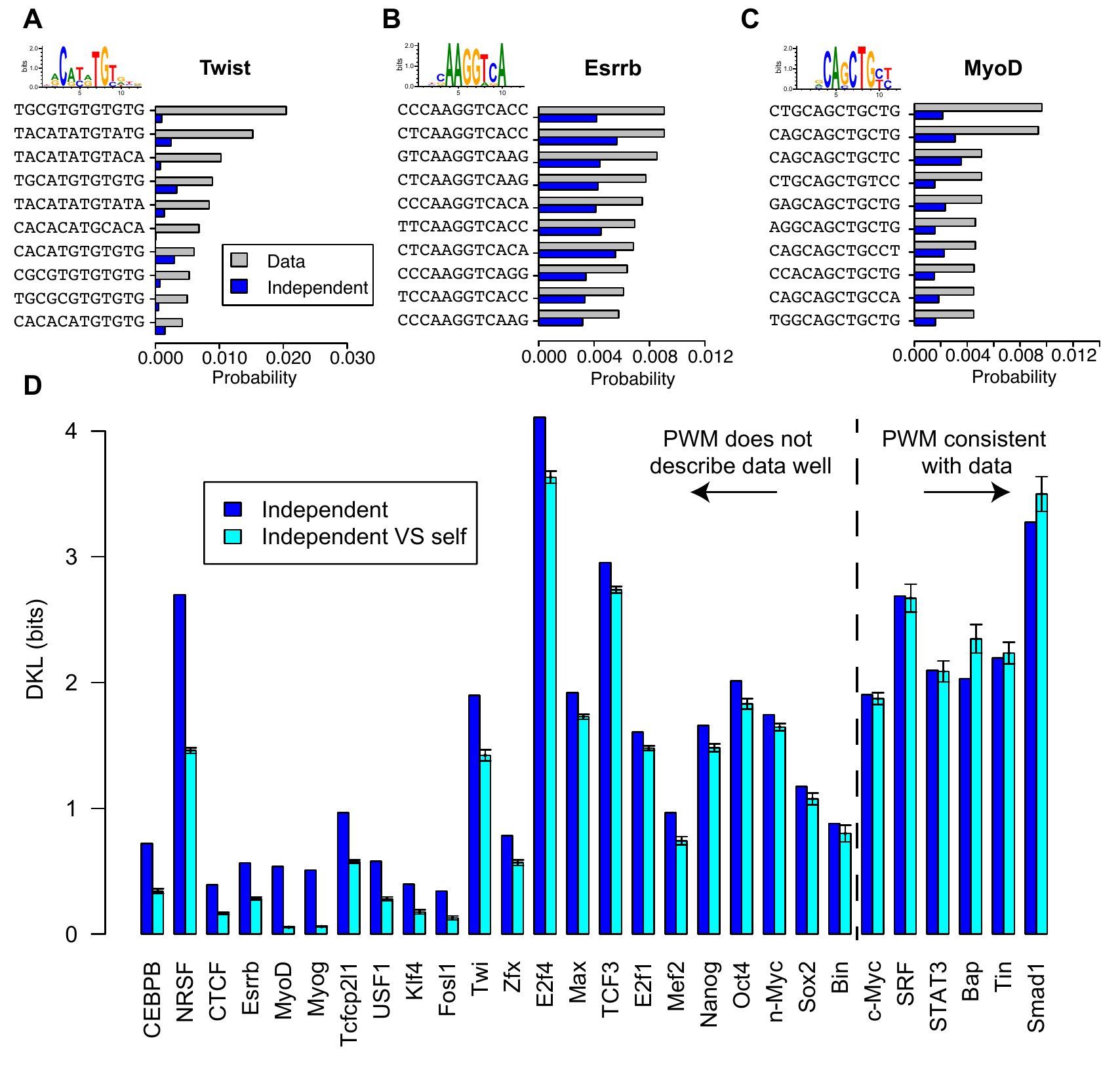}
\end{center}
\caption{
{\bf Observed TFBS frequencies are poorly predicted by a PWM model.}  The observed frequencies of the most represented binding site sequences for the TF Twist (A), Esrrb (B) and MyoD (C) are shown (gray bars) as well as the probabilities of these sequences as predicted by the PWM model (blue bars).
(D) Kullback-Leibler Divergence (DKL) between the observed probability distribution and the independent model distribution (blue). As a control we show the mean (cyan bars) along with two standard deviations of the DKL between the independent model and a finite sample drawn from it (see Methods). A discrepancy between the observed and predicted sequence probabilities is reported for 22 out of 28 factors. 
}
\label{fig:freqinde}
\end{figure*}

We first tested how well the usual PWM model reproduced the observed TFBS statistics, {\em i.e.} how well the frequencies of different TFBSs were retrieved by using only single nucleotide frequencies.
For this purpose, we used a collection  of ChIPseq data available from the literature  \cite{Zinzen:2009p760,Chen:2008p3738,encode12}, both from {\em  D. Melanogaster} and  from mouse embryonic stem cells (ESC) and a myogenic cell line (C2C12). The TFBSs are short $L$-mers (we take here $L=12$), which are determined in each few hundred nucleotides long ChIP-bound region with the help of a model of TF binding.
One important consequence and specific features of these data, is that the TFBS collection is not independent of the model used to describe it. 
Thus, in order to self-consistently determine the collection of binding sites for a given TF
from a collection of ChIPseq sequences, we iteratively refined the PWM together
with the collection of TFBSs in the ChIPseq data (see Figure~\ref{fig:workflow} and {\em Methods}). This process ensured that the frequency of different nucleotides at a given position in the considered ensemble of binding sites was exactly accounted by the PWM. We then enquired whether the probability of the different binding sequences in the collection agreed with that predicted by the PWM, as would be the case if the probabilities of observing nucleotides at different positions were independent. 
Figure \ref{fig:freqinde}
 displays the results for three different TFs, one from each of the three considered categories: Twi (Drosophila), Esrrb (mammals, ESC), and MyoD (mammals, C2C12).
For each factor, the ten most frequent sequences in the TFBS collection are shown. For comparison, Figure \ref{fig:freqinde} also displays the probabilities for these sequences as predicted by the PWM 
built from the TFBS collection. The independent PWM model strongly underestimates the probabilities of the most frequent sequences. Moreover, the PWM model
does not correctly predict the frequency order of the sequences and attributes comparable probabilities to these different sequences, in contrast to their observed frequencies.

The relative entropy or Kullback-Leibler divergence (DKL) is a general way to measure the difference between two probability distributions \cite{cover}.
In order to better quantify the differences between the observed binding sequence frequencies and the PWM frequencies, we computed the DKL between these distributions for all the considered TF,
as shown in Figure \ref{fig:freqinde}D. 
For each transcription factor T, part of the differences comes from the finite number N(T) of its observed binding sites. The results are thus compared for each factor T to DKLs between the PWM probabilities and frequencies obtained for artificial sequence samples of size N(T) generated with the same PWM probabilities. For most TFs (22 out of 28), the difference between the observed binding sequence frequencies and the PWM frequencies is significantly larger than expected from finite size sampling. In the following we focus on these 22 factors for which the PWM description
of the TFBSs needs to be be refined. It can be noted that
the 6 factors for which the PWM description appears satisfactory are predominantly those for which the smallest number of ChIP sequences is available (see Table 1 and Figure \ref{corrseqnumbimprov}).

\subsection*{Pairwise interactions in the binding energy improve the TFBS description}

The discrepancy between the observed statistics of TFBSs and the statistics predicted by the PWM model calls for a re-evaluation of the PWM main hypothesis, namely the independence of bound nucleotides. As recalled above, the inverse problem of devising interaction parameters from observed frequencies of ``words'' has been recently studied in different contexts. It has been proposed to include systematically pairwise correlations between the ``letters'' comprising the words to refine the independent letter description. In the case of a two-letter alphabet, the obtained model 
is equivalent to the classical Ising model
of statistical mechanics\cite{baxter}. In the present case, the 4-nucleotide alphabet (A,C,G,T) leads to a model equivalent to 
the so-called inhomogeneous Potts model \cite{baxter} (hereafter called pairwise interaction model), a generalization of the Ising model to the case where spins assume $q$ values and their fields and interaction parameters depend on the sites considered. In this analogy, nucleotides are spins with $q=4$ colors.

In practice, the probability of observing a given word $(s_1...s_L)$ in the dataset is expressed as $P[s_1...s_L]=(1/\mathcal{Z})\exp(-\mathcal{H}[s_1...s_L])$, where $\mathcal{Z}$ is a normalization constant. $\mathcal{H}$ is formally equivalent to a Hamiltonian in the language of statistical mechanics, and reads:
\begin{equation}
\begin{split}
\mathcal{H}[s_1...s_L] =& - \sum_{i=1}^{L} h_i(s_i) - \sum_{i=1}^{L} \sum_{j<i} J_{i,j}(s_i,s_j), \\ &\qquad s_i \in \{A,C,G,T\}
\end{split}
\label{pottsh}
\end{equation}
The ``magnetic fields'' $h_i$ at each site $i$, along with the interaction parameters $J_{ij}$ between nucleotides at positions $i$ and $j$, are computed so as to reproduce the frequency of nucleotide usage at each position in the TFBS as well as the pairwise correlations between nucleotides at different positions 
(see {\em Methods}). In principle, the number of parameters in the model is sufficient to reproduce the observed values of all pairwise correlations between nucleotides.
This however would result in over-fitting the finite-size data with an irrealistically large number of parameters. Therefore, to obtain the model
parameters we instead maximized the
likelihood that the data was generated by the model with a penalty proportional to the numbers of parameters involved, as provided by
the Bayesian Information Criterion (BIC) \cite{bishop}. Similarly to the procedure followed for the PWM, the pairwise interaction model and the collection of TFBSs for a given factor were iteratively refined together, as schematized in Figure \ref{fig:workflow}.

\begin{figure*}[ht]
\begin{center}
\includegraphics[width=.7\linewidth]{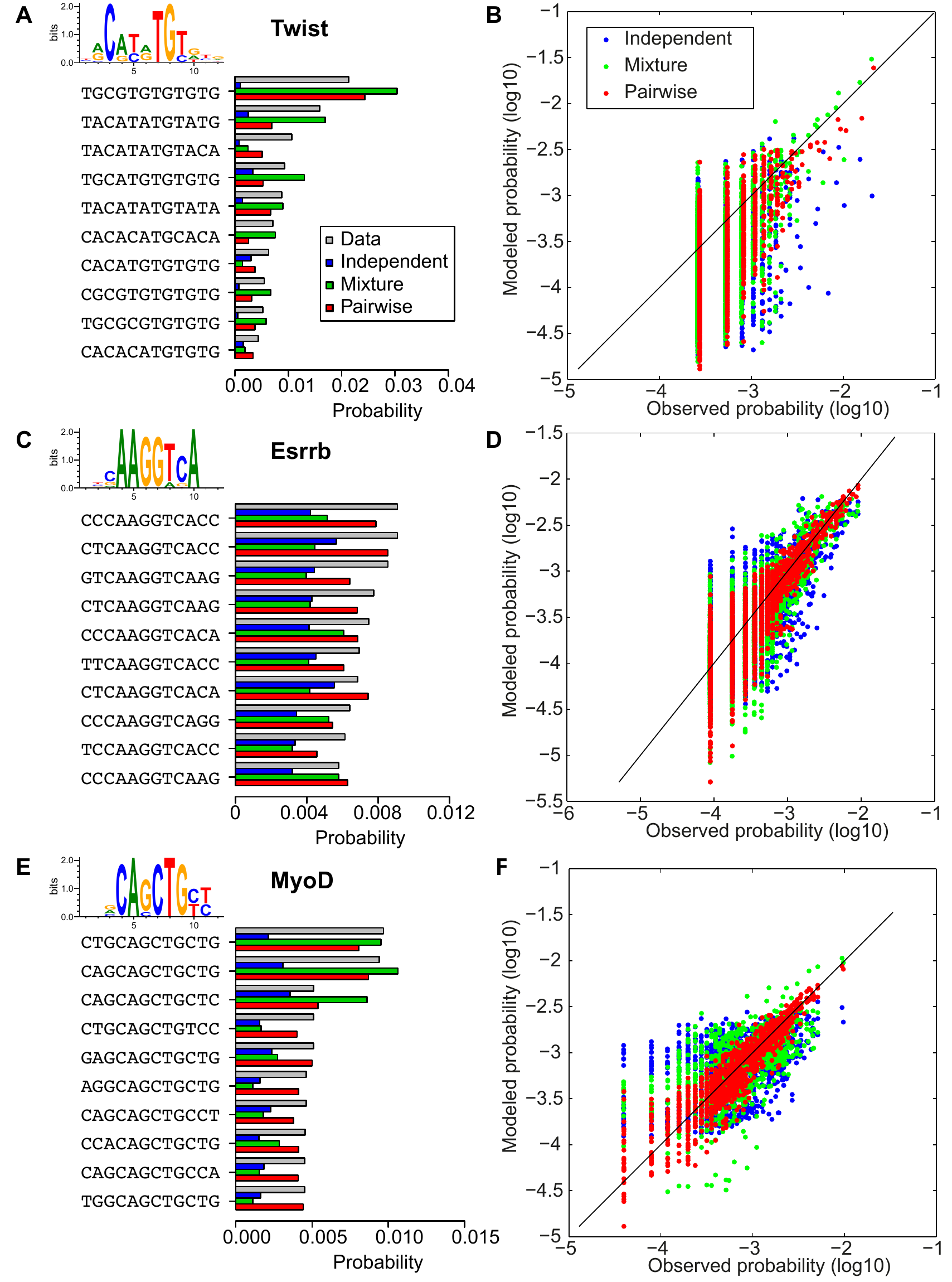}
\end{center}
\caption{
{\bf Models with correlations improve  TFBS statistics prediction.} The observed frequencies (gray bars) of the most represented TFBSs  for Twist (A), Esrrb (B) and MyoD (C) TFs, are shown
together with the probabilities of these sequences predicted by the independent energy model (blue bars), the pairwise model taking into account interactions between nucleotides (red bars), and the K-means mixture model (green bars). (B,D,F) show the comparison between  frequencies for all binding sequences and predicted sequence probabilities for the three models (same color code). 
The probability predictions of the pairwise model and to a lesser extent of the mixture model are in much better agreement with the observed frequencies than those of the PWM model.
}
\label{fig:probamodels}
\end{figure*}

Figure \ref{fig:probamodels} shows the improvement in the description of TFBS statistics when using the final pairwise interaction model, for the three factors chosen for illustrative purposes. Where the independent model failed at reproducing the strong amplitude and non-linear decrease in the frequencies of the most over-represented TFBSs, the pairwise interaction model provides a substantial improvement in reproducing the observed statistics. The improvement is most apparent when comparing the  frequencies of the ten most observed TFBSs between the model and the ChIPseq data (Figure \ref{fig:probamodels} A, C, E), and is further shown by the statistics of the full collection of TFBSs (Figure \ref{fig:probamodels} B, D, F).

\subsection*{The pairwise model ranks binding sites differently from the PWM}
Precise predictions of TFBSs are one important output of ChIPseq data. 
Moreover, they condition further validation experiments such as gel mobility shift assays or mutageneses. We therefore found it worth
assessing the difference in TFBS predictions between pairwise and independent models.

\begin{figure*}[ht]
\begin{center}
\includegraphics[width=0.7\linewidth]{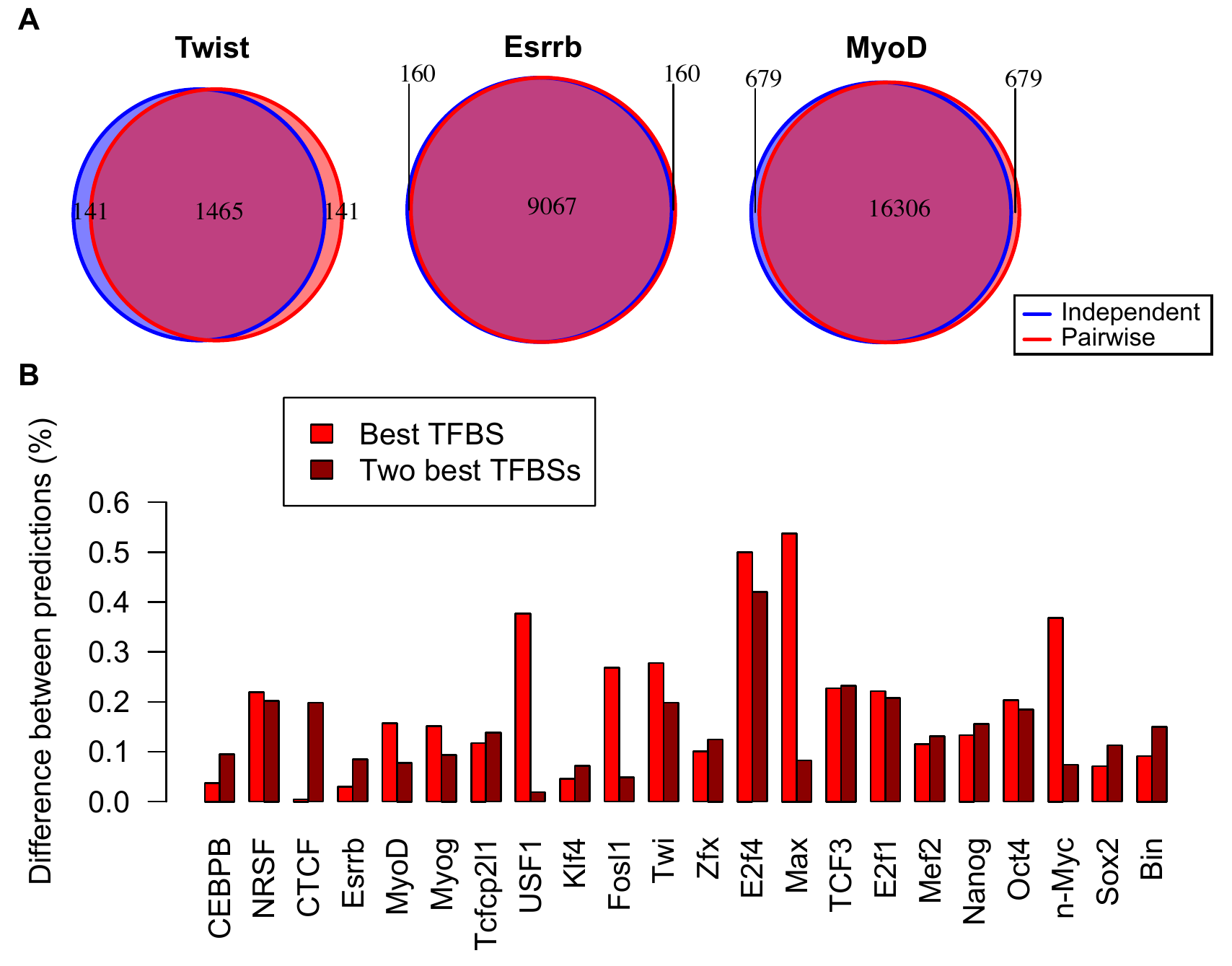}
\end{center}
\caption{
{\bf Overlap between predicted sites.} 
(A) Venn diagrams showing the overlap between the ChIP predicted by the independent (blue) and pairwise (red) models. 
(B) Difference (one minus the proportion of shared sites) 
between the best sites predicted by pairwise and PWM models on ChIPseq peaks (light red), and the same quantity when including the next best predicted sites on each peak (dark red). In several cases ({\em e.g.}  Fosl1, Max, n-Myc, Srf, Stat3, Usf1), the difference between predicted sites
is much smaller when the two best sites are considered, 
indicating that the pairwise model and the PWM model rank differently the two best sites in ChIP peaks with multiple bound sites.
}
\label{fig:overlap}
\end{figure*}

First, we 
compared the set of ChIP sequences
retrieved by the independent and pairwise models
model 
at the cutoff of $50\%$ TPR (True Positive Rate) used in the learning scheme, as shown in Figure \ref{fig:overlap}A. The non overlapping set of ChIPseq sequences ({\em i.e.} sequences that were picked by one model but not by the other) was found to range from a few percent for TF like Esrrb, up
to about 15 \% for Twist. Thus, even when stemming from the same ChIPseq data, the two models can be learnt from significantly distinct set of sites.

Second, 
using the set of ChIPseq peaks on which the pairwise model was learned, we looked for the best predicted sites on each ChIPseq bound fragment using both the pairwise and PWM models (Figure \ref{fig:overlap}B).

 The overlap was found to be about $80\%$ on average. 
The overlap between the sets comprising the two best TFBSs of each ChIPseq was also computed. This resulted in an overlap increase or decrease between the prediction of the two models
depending on the average of number of binding sites per retrieved ChIPseq fragment. In a few cases ({\em e.g} CTCF, Esrrb), the inclusion of the second
best TFBS increased the difference between the two models. This generally happened when the ChIPseq fragments were retrieved with typically a single TFBS above threshold ({\em e.g.} for Essrb the TFBS specificity was fixed to retrieve 50\% of 18453 ChIPseq and about 11000 fragments where found by the two models---see Table I). In these cases, the low specificity TFBSs tended to differ more between the two models than the very specific ones.
In several other cases ({\em e.g} for Fosl1, Max, n-Myc, USF1), the inclusion of the second best predicted binding sites (Figure \ref{fig:overlap}B) greatly increased the overlap between the two model predictions. This corresponded to cases for which the retrieved fragments contained on average two of more TFBSs about the specificity threshold (Table I). This showed that for these cases the prediction difference between the two models arose predominantly from a different ranking of the best TFBSs. 

In conclusion, the TFBS predictions made by the two models can differ significantly both in the rank of ChIPseq fragments and in the rank of binding sites on these fragments.

\subsection*{Comparison with a PWM-mixture model}

\begin{figure*}[!ht]
\begin{center}
\includegraphics[width=.7\linewidth]{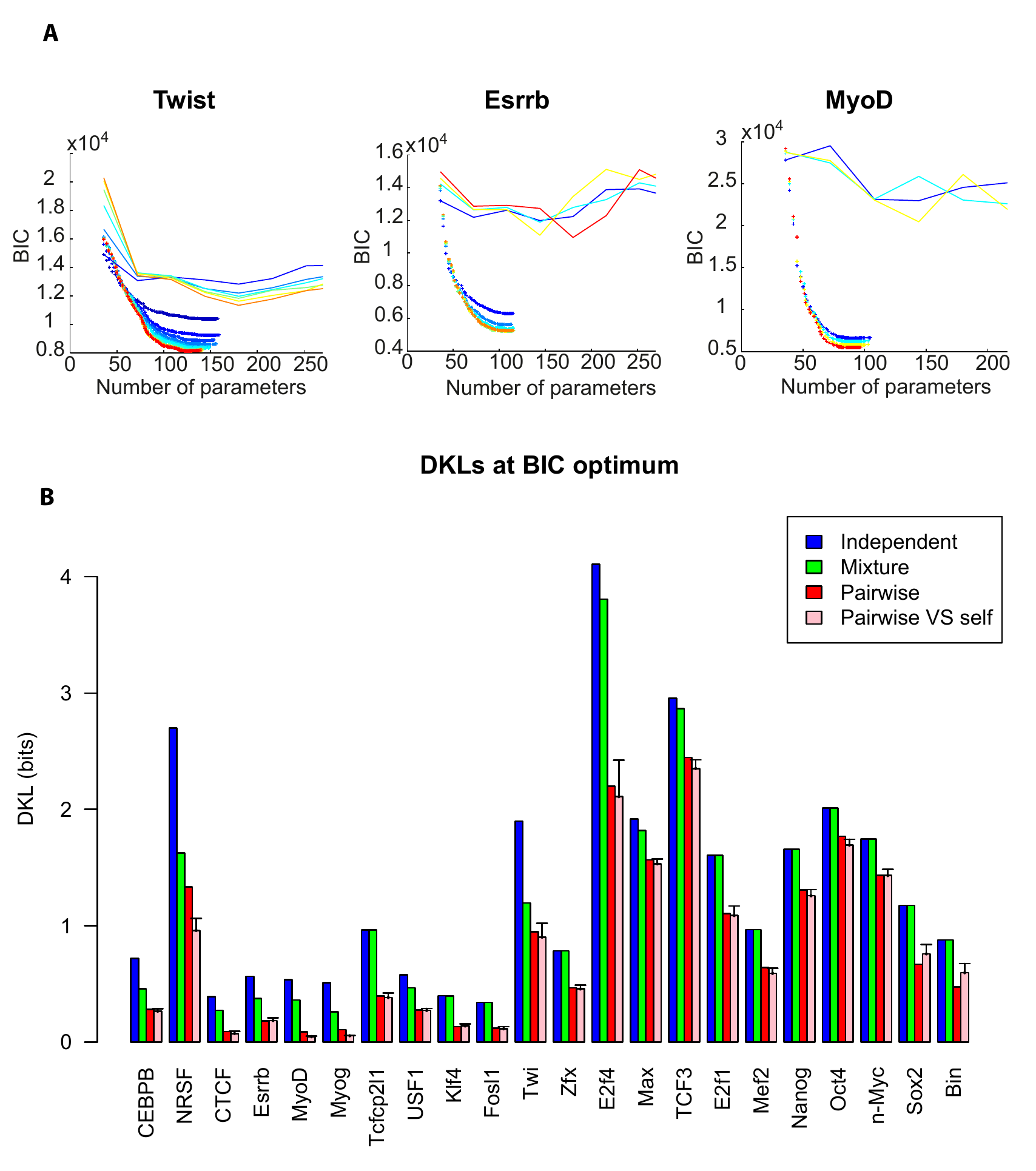}
\end{center}
\caption{ 
{\bf Model selection.}  (A) Minimisation of the Bayesian information criterion (BIC, see {\em Methods}) is used to select the optimal number of model parameters  and avoid over-fitting the training set. The evolution of the BIC is shown for the pairwise model (crosses) and the PWM-mixture model (lines). Colors from dark blue to red indicate the number of interations (see Fig.\ref{fig:workflow}).\\
(B)  Kullback-Leibler divergences (DKL) between the independent, K-means and pairwise distributions and the observed distribution for the different TFs, for the BIC optimal parameters. We also show the DKL of the pairwise model with a finite-size sample of sequences drawn from it (pink, see {\em Methods}). Error bars represent two standard deviations.
}
\label{fig:bic}
\end{figure*}

When described by a PWM, the binding energies of a TF for different nucleotides sequences form a simple energy well with a single minimum at a preferred consensus sequence. 
Some authors have instead analyzed the binding specificity of transcription factors by introducing multiple
preferred sequences \cite{Cao:2010fk,Heinz:2010fk}. A model of this type that naturally generalizes the PWM description consists of using multiple PWMs 
\cite{barash2003mod}.
We found it interesting to investigate this approach based on a mixture of PWMs and compare it with the pairwise interaction model 
to get some insights into
potentially important high-order correlations that would not be captured by the pairwise model. As precisely described in {\em Methods}, an initial mixture of
K PWMs was generated by grouping into $K$ clusters the TFBS data for a given TF. Similarly to the pairwise interactions, the number of clusters $K$ was constrained,
to avoid over-fitting, by penalizing the corresponding model score using the BIC.
For a given TF, the PWM mixture and
the collection of TFBSs in the ChIPSeq data were refined iteratively until convergence, usually reached after $10$ iterations.
The results are shown in Figure \ref{fig:bic}A for the three representative factors, Twi, Esrrb and MyoD.

The best description of Twi ChIPSeq data is, for instance, provided
by a mixture of 5 PWMs, which corresponds to 184 
independent parameters. The mixture model yields a significant improvement when compared to the single-PWM model for Twi, and milder ones for Essrb and MyoD. In the three cases however, it proves inferior to the pairwise model.

More generally, Figure \ref{fig:bic}B shows the performances of the different models for all studied TFs using the Kullback-Leibler Divergence or DKL between the data distribution $P(s)$ and the models distributions $P_m(s)$.
On the one hand, the mixture model improves the description of the binding data for 12 out of 27 TFs as compared to the single PWM model. The mixture model gives in particular strong improvements in the cases for which the binding sites have a palindromic structure (eg Twi, MyoD, Myog, Max, USF1). This feature often stems from the fact that the TF binds DNA as a dimer, 
which could give some concreteness to the mixture model:
the recruitment of different partners by bHLH factors like MyoD or Myog could indeed
lead to a mixture of TFs binding the same sites. On the other hand, the pairwise model clearly outperforms the other models in all cases studied. 

As in the PWM case, the finite size of the datasets leads us to expect 
fluctuations in the estimation of the DKL. In order to assess the magnitude of these finite-size fluctuations, 
we computed the average DKL between the best-fitting (pairwise) model and a finite-size artificial sample drawn from its own distribution, as shown
 in Figure \ref{fig:bic}B. Values of this DKL that are larger than the one obtained with the real dataset are indicative
 of overfitting, while the opposite case would suggest that the model is incomplete.
In all cases, however, the DKL obtained with this control procedure was within error bars of the value computed with respect to the observed sample, with the exception of NRSF, MyoD, and Myog, as seen in Figure \ref{fig:bic}B. 
Thus, the pairwise model is generally the best possible model, 
insofar as the available dataset allows us to probe.

\subsection*{The metastable states of the pairwise interaction model}
In order to more directly relate the pairwise interaction and the mixture models, it is useful to consider the energy landscape of the pairwise interaction model
in the space of all possible TFBSs. By contrast with the simple, single-minimum energy well of the PWM model, the pairwise interaction model has multiple
metastable energy minima. The energy landscape of the pairwise interaction model can thus be seen as a collection of energy wells, each centered on its metastable
energy minimum. The span of the different energy wells in sequence space can be precisely defined as the basins of attraction of the different metastable minima in an energy minimizing procedure (see {\em Methods}). This allows one to associate each observed TFBS to a particular energy minimum. This defines basins of attraction that are used to build
representative PWMs for each metastable minimum together with a weight---the number of sequences in the basin of attraction---for this energy minimum.
We compared each metastable minimum to the PWMs of the mixture model,
by calculating the DKL between the PWM computed from the sequences in its basin of attraction and the PWMs of the mixture model. This gave an effective distance which allowed us to associate each metastable state to the nearest PWM of the mixture model.

\begin{figure*}[ht]
\begin{center}
\includegraphics[width=\linewidth]{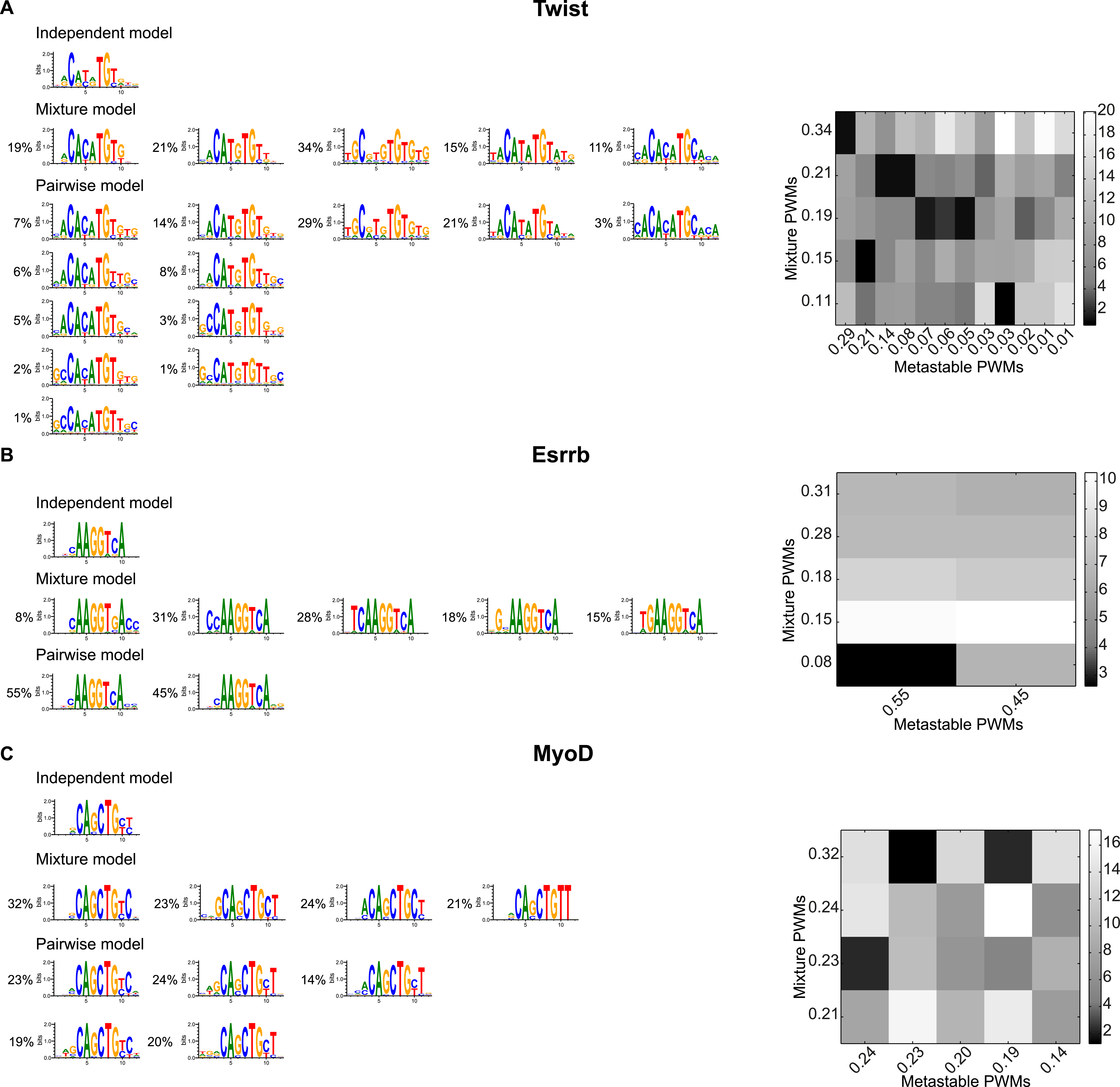}
\end{center}
\caption{
{\bf Metastable states.}  The DNA sequence variety described by each model is illustrated using weblogos \cite{Crooks:2004p3737}. Shown are PWMs built from all sites, from the PWM-mixture model, and from the basins of attraction of the pairwise interaction model for Twist (A), Esrrb (B), and MyoD (C). The metastable PWMs are grouped under the mixture PWMs with smallest distance (measured by DKL, in bits).  Heatmaps showing the DKLs between metastable PWMs and mixture PWMs are displayed on the right for each factor (minimal DKLs are in black). The proportions of sites used for each logo are also indicated and serve to denote the corresponding PWM.
}
\label{fig:metastable}
\end{figure*}

Using this procedure, we computed the set of PWMs and weights corresponding to the 27 considered TF pairwise interaction models.
Examples are shown in Figure \ref{fig:metastable}. In the case of Twi, the PWMs of the pairwise model (``metastable PWMs'') can be clearly associated to the $K=5$ PWMs of the mixture model. For MyoD, three of the 5 ``metastable PWMs'' can be clearly assigned to PWMs of the mixture model. The other two have a more spread out representation. The case of Esrrb is
similar with one ``metastable PWM'' in clear correspondence with one PWM of the mixture model, and the other one less clearly so.
The correspondence between the two models is shown in
Figure 
\ref{figS2-usf1} for the other TFs for which the mixture model uses more than a single PWM. This representation allows one to identify some features captured by the pairwise model. For example, in the case of Twist, most of the correlations are coming from the two nucleotides at the center of the motif, which take mainly $3$ values among the $16$ possible: CA,TG and TA. In the 
case of MyoD, the representation makes apparent 
the interdependencies between the two nucleotides following the core E-Box 
 motif, and the restriction to the three main cases of 
CT, TC and TT.

\subsection*{Properties of the pairwise interactions}

The computation of
the interaction parameters
allows an
analysis of some of their properties. In particular, it is interesting to quantify their strengths and measure
 the typical distance between interacting nucleotides. We address these two questions in turn.

\begin{figure*}[ht]
\begin{center}
\includegraphics[width=.8\linewidth]{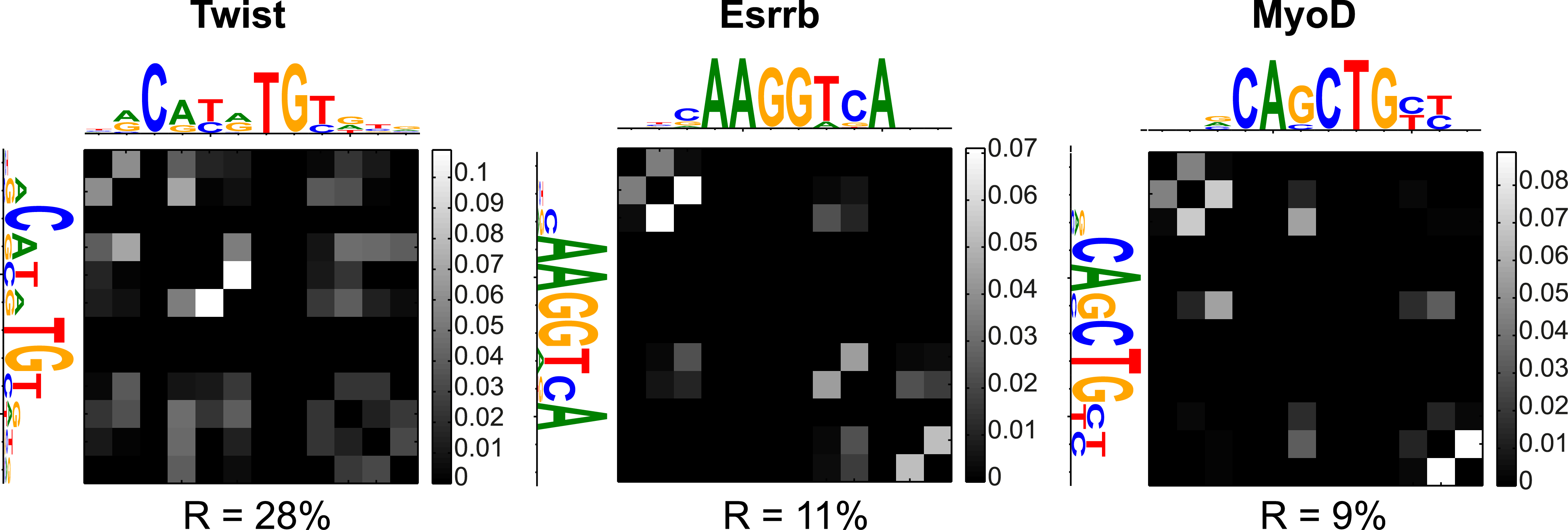}
\end{center}
\caption{
{\bf  Nucleotide pair interactions.}  Heat maps showing the values of the Normalized Direct Information between pairs of nucleotides. The matrix is symmetric by definition. PWMs are shown on the side for better visualization of the interacting nucleotides. The participation ratio R is indicated below each heat map.
}
\label{fig:dirinfo}
\end{figure*}

The concept of Direct Information was previously introduced to predict 
contacts between residues from large-scale correlation data in protein families \cite{Weigt:2009p2811}. We used it
here to measure the strength of the pairwise interaction between two nucleotides. Using the previously generated interaction parameters from the pairwise model, we built the Normalized Direct Information (NDI), a quantity which varies from 0 for non-existing interactions, to 1 when interactions are so strong that knowing the amino acid identity at one position entirely determines the amino acid identity at the other position
(see {\em Methods}).
 Heatmaps displaying the results for the representative Twist, Esrrb and MyoD factors are shown in Figure \ref{fig:dirinfo} and in Figures
 \ref{fig:dirinfoSI2} for the other factors. An important observation is that the direct information between 
 different nucleotides is rather weak---usually smaller than $10\%$---but substantially larger than the direct interaction between nucleotides in the surrounding background (1-3\%, see Figure \ref{backcor}). It is interesting to note that such weak pairwise interactions give rise to a substantial improvement in the description of TFBS statistics, similarly to what was previously found in a different context
 \cite{Schneidman:2006p2295}. 
 The pairwise interactions are furthermore observed in Figure~\ref{fig:dirinfo} to be concentrated on a small subset of all possible interactions. This can be made quantitative by computing the  
Participation Ratio of the interaction weights,
 an indicator of the fraction of 
 pairwise interactions that accounts for
  the observed Direct Information (see Methods). Here, typical values of $10-20\%$ were found (Figure  \ref{fig:dirinfo} and Table \ref{tab:partratio}),
  showing that the interactions tend to be concentrated on a few
   nucleotide pairs.

\begin{table}[ht]
\caption{
\bf{Participation Ratios}}
\begin{tabular}{|c|c|}
\hline
Name & Part. Ratio \\ \hline 
 Bin & $0.11$ \\ \hline
Mef2 & $0.19$ \\ \hline
Twi & $0.28$ \\ \hline
\hline
E2f1 & $0.13$ \\ \hline
Esrrb & $0.11$ \\ \hline
Klf4 & $0.16$ \\ \hline
Nanog & $0.10$ \\ \hline
n-Myc & $0.09$ \\ \hline
Oct4 & $0.24$ \\ \hline
Sox2 & $0.12$ \\ \hline
Tcfcp2l1 & $0.12$ \\ \hline
Zfx & $0.10$ \\ \hline
\hline
CEBPB & $0.05$ \\ \hline
CTCF & $0.23$ \\ \hline
E2f4 & $0.14$ \\ \hline
Fosl1 & $0.09$ \\ \hline
Max & $0.18$ \\ \hline
MyoD & $0.09$ \\ \hline
Myog & $0.09$ \\ \hline
NRSF & $0.27$ \\ \hline
TCF3 & $0.19$ \\ \hline
USF1 & $0.07$ \\ \hline
\end{tabular}
\label{tab:partratio}
 \end{table}

\begin{figure*}[ht]
\begin{center}
\includegraphics[width=.7\linewidth]{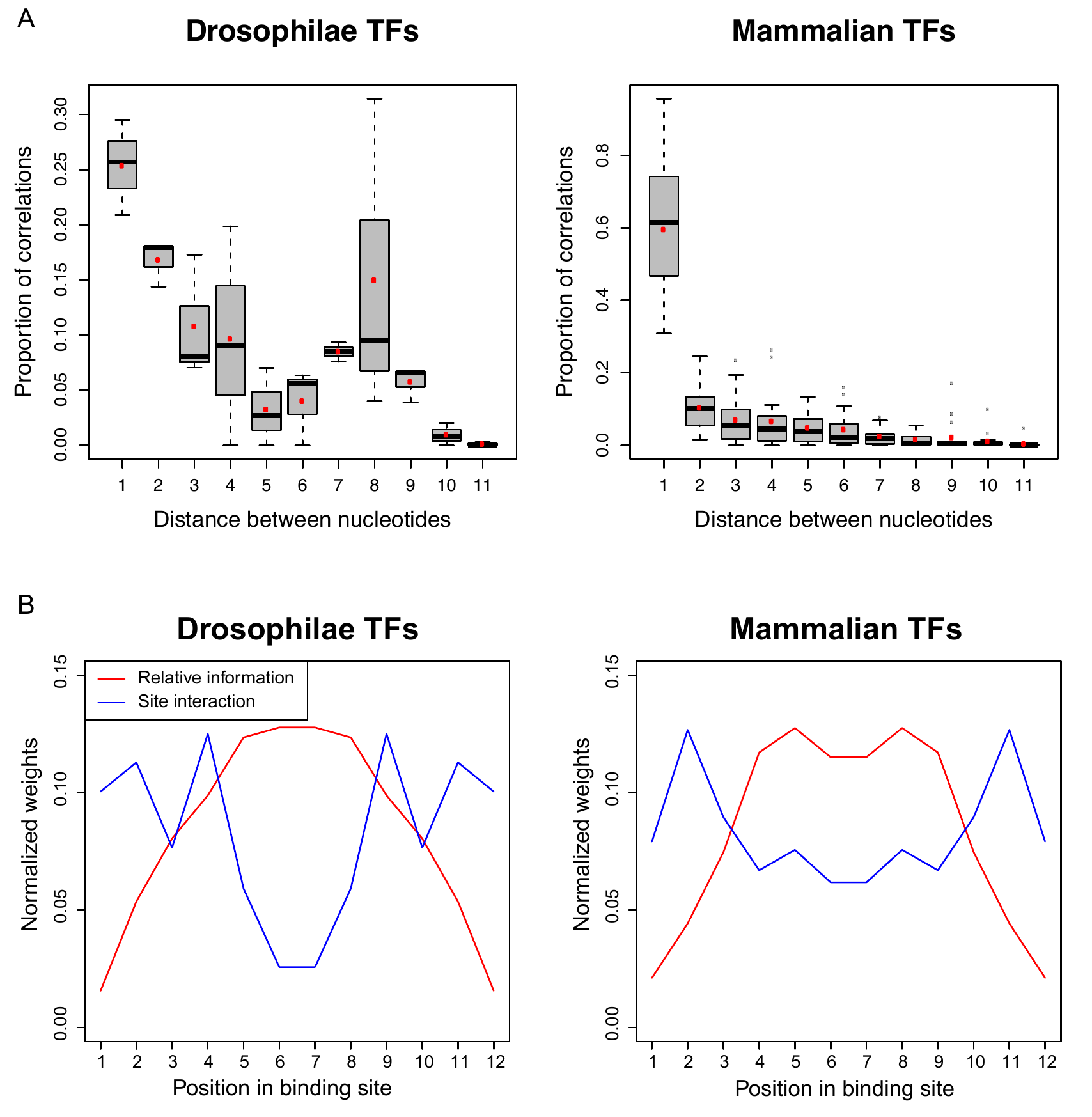}
\end{center}
\caption{
{\bf Properties of the pair interactions.} (A) Distances between interacting nucleotides. The box plots show the relative importance of the Normalized Direct Information as a function of the distance between interacting nucleotides. Red dots denote average values. (B)  Sum of normalized direct informations in the TFBSs at a given position, averaged over all considered factors (blue line). The average site information content relative to background as a function of position is also shown (red line). In both quantities, the average over the two TFBS orientations has been taken. 
}
\label{fig:corrdist}
\end{figure*}

The interaction weights can also be used to measure the typical distance between interacting nucleotides.
To that purpose, we computed the relative weight of the Direct Information as a function of the distance between nucleotides (see Methods). Figure \ref{fig:corrdist} A shows box plots that summarize the results for the considered Drosophilae and mammalian TFs. Both plots show a clear bias towards nearest-neighbor interactions with a strong peak at $d=1$, and a rapid decrease for $d\geq2$. Finally, the dominant pair interactions 
are on average located in the flanking regions of the
BS in clear anti-correlation with the most informative nucleotides which are on average in the central region (Figure \ref{fig:corrdist} B). These observations for TF binding {\em in vivo} agree with similar ones made from a large recent analysis of TF binding {\em in vitro} \cite{jolmataipale13cell}. The fact  that for pair correlations to be important,  nucleotide variation at a given location is required, may be one way to rationalize them.

\subsection*{Alternative representation of interactions by Hopfield patterns}
\begin{figure*}[ht]
\begin{center}
\includegraphics[width=0.7\linewidth]{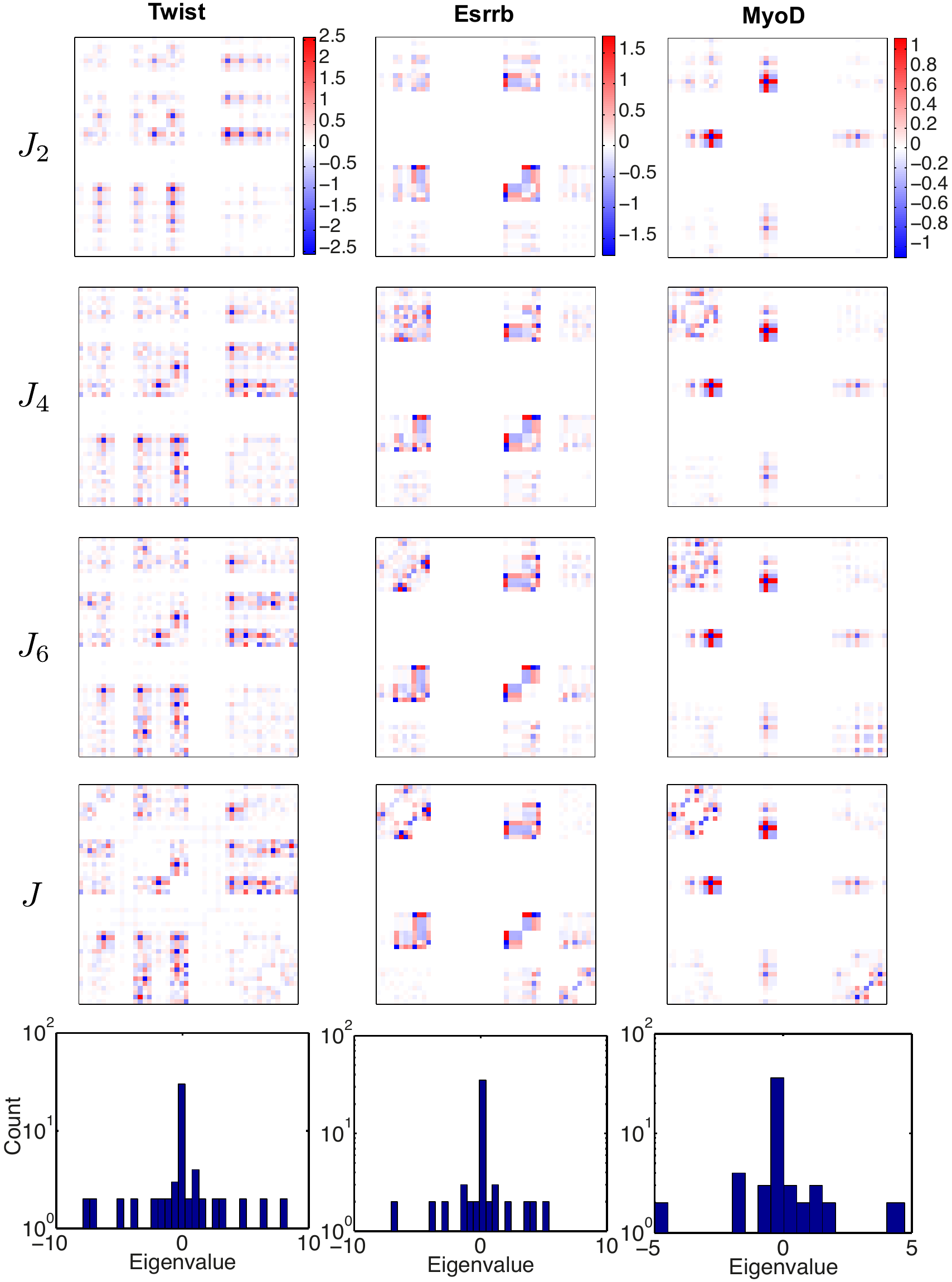}
\end{center}
\caption{
{\bf Representation of interactions by Hopfield patterns.}
The full interaction matrix $J$ is approximated by a matrix $J_p$ built from the $p$ Hopfield patterns with highest eigenvalue moduli. We show $J_2$, $J_4$, $J_6$ and the full matrix $J$ in the basis $(i,b)$ with $i=\{1,...,12\}$ and $b=$\{A,C,G,T\}. Color bars are shown on the first row for each factor. For MyoD, the correspondence between successive pairs of patterns and distinct interaction domains is seen particularly clearly. In all cases the full interaction matrix is already well approximated by $J_6$. 
}
\label{fig:pattern}
\end{figure*}

Using a simple binary description of neurons, JJ Hopfield suggested, in a classic piece of work \cite{hopfield82pnas}, that neural memories could be attractors corresponding to patterns arising from pair interactions between neurons.
These interaction patterns can be computed in the present case. They offer an
alternative way to analyze the patterns of correlation from the pair-interactions between positions, as  already proposed in a mean-field context in \cite{cocco11pre}.
Because the matrix of interactions $J_{ij}$ is symmetric, it can be diagonalized in an orthonormal basis of eigenvectors  $\mathrm{\xi}^k$, the Hopfield patterns in the present case,
with corresponding real eigenvalues $\lambda_k$. These orthonormal eigenvectors correspond to  the Hopfield patterns in the present case.
The Potts energy (Eq.~(\ref{pottsh})) for a binding sequence $s_1\cdots s_L$ can be  rewritten in terms of the Hopfield patterns as (see Methods):
\eqn{
\label{eq:hopfield}
\mathcal{H}=-\sum_i h_i(s_i) - \frac{1}{2}\sum_{k=1}^{4L} \lambda_k \left(\sum_{i=1}^L \xi_{i}^k(s_i)\right)^2.
}
Although here the presence of the diagonal $h$ term prevents the patterns to be metastable energy states, they can still be useful to analyze the interaction matrix.
This spectral decomposition of the interaction matrix is also similar in spirit to a principal component analysis, and even equivalent in the case of Gaussian variable.
One can thus wonder how many patterns are needed to well approximate  the full matrix of interactions $J$. To address this question, one can rank the eigenvalues $ \lambda_k$ in order of decreasing moduli and note $J_p$ the restriction of the interaction matrix generated by the first $p$ eigenvalues and their associated patterns. 
The full interaction matrix  naturally corresponds to  $J_{48}$. Approximate interaction matrices obtained by keeping different numbers of dominant patterns are shown in Figure \ref{fig:pattern} for the three considered representative factors. Pairs of successive patterns appear to provide the main interaction domains in this representation, as is particularly clear in the case of MyoD. One can see in Figure \ref{fig:pattern} that  $J_6$ already closely approximates the full  interaction matrix, a reflection in the present representation, that the important interactions are concentrated on a few links between pairs of nucleotides.

\section*{Discussion}

The availability of ChIPseq data for many TFs has led us to revisit the question of nucleotide correlations in TFBSs. In order to perform a fully consistent analysis of this type of data, we have developed
a workflow in which the TFBS collection and the model describing them are simultaneously obtained and refined together. 
We have found that when a sufficiently large number of TFBSs is available, the PWM description does not account well for their statistics. The general presence of correlations that follows from this finding, agrees with previous reports for particular transcription factors 
 \cite{Man:2001p12827,Bulyk:2002p12734,Cao:2010fk} and with the conclusions of  large scale {\em in vitro} TF binding studies \cite{Badis:2009p12808,jolmataipale13cell}.  
 
 In order to refine the PWM description, we have analyzed a model with pairwise interactions \cite{zhaostormo12gen}, and a PWM mixture model \cite{barash2003mod}. Data overfitting is a concern for multi-parameter models and has been addressed by putting a penalty on the parameter number  using the BIC.
 While the mixture-model improved in some cases the PWM description, especially for palindromic binding sites, a much more significant and general improvement was found with the pairwise interaction model.  The success of  the pairwise interaction model  agrees  with the results of its recent application (however, without the BIC)
to high-throughput {\em in vitro}  binding data \cite{zhaostormo12gen}.  
It moreover shows that, at least in the case we considered, pairwise interactions are sufficient to account for higher-order correlations, and that an explicit
description like the one provided by the PWM-mixture model is not necessary. 
For example, for Essrb, metastable states arising from nearest-neighbor interactions reproduce a triplet of flanking nucleotides with a variable spacer from the core motif (Figure \ref{figspacer}).

Our detailed analysis of the obtained interaction models for different TFs shows that the weights of pairwise interactions are generally weak. The most important are only about 10 \% of the PWM weights, but significantly above the interaction weights in the surrounding background  DNA (of the order 1-3\% by the
same measure). Nonetheless, collectively these interactions  significantly improve the model description of the TF binding data as found in other examples \cite{Schneidman:2006p2295}.

 We have here obtained the pairwise interaction models based on the principle of maximum entropy, constrained to account for the pair-correlations measured in the data. This approach has already been followed
in a variety of biological contexts, from populations of spiking neurons \cite{Schneidman:2006p2295, Shlens:2006p1442} to protein sequences \cite{Weigt:2009p3341} to bird flocks \cite{Bialek:2012p12539}. 
An interesting feature of these  interaction models is their non-convexity, which allows for the existence of many local maxima in the probability distribution of sequences, or local minima of energy. This was noted for repertoires of antibodies in a single individual \cite{Mora:2010p5398}, where many of these local states were observed and suggested as possible signatures of past infections. In a very different context, local probability maxima in the probability distribution of retinal spiking patterns was reported and linked to error-correcting properties of the visual system \cite{Tkacik:2006p1289}. In the  present case of TFBSs, these local mimima reflect the multiplicity of binding solutions  and resemble the individual PWMs  of the mixture model. Pairwise interaction models  thus somehow incorporate models of multiple PWMs while outperforming them.
 
 The previously considered  case of protein sequences shares many similarities to the statistics of TFBSs, since correlations in protein sequences as in TFBSs reflect both structural and functional contraints. 
In proteins families, correlations are usually interpreted as resulting from the co-evolution of residues interacting with each other in the protein structure. These effects are hard to distinguish from phylogenic correlations or other observational biases. 
Nonetheless, the inference of interaction models from data was successfully  used to predict physical contacts between amino-acids in the tertiary structure \cite{Morcos:2011p12728}, and to aid molecular dynamics simulations in predicting protein structure \cite{Sukowska:2012p12846,Hopf:2012p12848,Marks:2011p12847}. 
In the case of TFBSs, comparison between {\em in vitro}\cite{Badis:2009p12808,jolmataipale13cell} and {\em in vivo} binding data may help to disentangle the different possible origins of the found correlations and seems  worth  pursuing. 
It appears similarly interesting to study how much of the found pair correlations can be explained on the basis of structural data.
Finally, the role of nucleotide interaction in TFBS evolution \cite{lassig07} should be considered and could improve the reconstruction of TFBSs from multi-species comparison \cite{moses2004monkey,siddharthan2005phylogibbs,Rouault:2010fk}. 

Independently of these future prospects, we have found that the TFBSs predicted from  ChIP-seq data significantly depended on the model used to extract them. Since the pairwise interaction model and the developed workflow significantly improve TFBS description and require a modest computational effort, they should prove worthy tools  in  future data analyses.

\begin{table*}[!ht]
\caption{
\bf{Information about the TFs}}
\begin{tabular}{|c|c|c|c|c|c|c|c|}
\hline
Name & $N_{\text{chip}}$ & $\Delta_{\text{inde-mixture}}$ & $\Delta_{\text{inde-pairwise}}$ & $\Delta_{\text{mixture-pairwise}}$ & $N_{\text{inde}}$ & $N_{\text{mixture}}$ & $N_{\text{pairwise}}$ \\ \hline 
 Bap & $678$ & $0$ & $12$ & $12$ & $2205$ & $2208$ & $2117$ \\ \hline
Bin & $1857$ & $2$ & $80$ & $81$ & $1300$ & $1298$ & $1228$ \\ \hline
Mef2 & $4545$ & $0$ & $161$ & $161$ & $3681$ & $3681$ & $3665$ \\ \hline
Tin & $1791$ & $0$ & $40$ & $40$ & $1333$ & $1333$ & $1310$ \\ \hline
Twi & $3211$ & $182$ & $141$ & $128$ & $3810$ & $3862$ & $3722$ \\ \hline
\hline
c-Myc & $3038$ & $0$ & $95$ & $95$ & $2996$ & $2996$ & $2920$ \\ \hline
E2f1 & $17367$ & $0$ & $877$ & $877$ & $16625$ & $16625$ & $14915$ \\ \hline
Esrrb & $18453$ & $172$ & $160$ & $167$ & $11243$ & $11333$ & $11275$ \\ \hline
Klf4 & $9404$ & $0$ & $97$ & $97$ & $5912$ & $5912$ & $5913$ \\ \hline
Nanog & $8022$ & $0$ & $111$ & $111$ & $6196$ & $6196$ & $6224$ \\ \hline
n-Myc & $6367$ & $0$ & $54$ & $54$ & $6981$ & $6981$ & $6954$ \\ \hline
Oct4 & $3147$ & $0$ & $74$ & $74$ & $3187$ & $3187$ & $3079$ \\ \hline
Smad1 & $907$ & $0$ & $24$ & $24$ & $690$ & $690$ & $667$ \\ \hline
Sox2 & $3523$ & $0$ & $95$ & $95$ & $2306$ & $2306$ & $2293$ \\ \hline
STAT3 & $2099$ & $54$ & $58$ & $62$ & $2308$ & $2264$ & $2231$ \\ \hline
Tcfcp2l1 & $22406$ & $0$ & $418$ & $418$ & $16691$ & $16691$ & $16649$ \\ \hline
Zfx & $9152$ & $0$ & $203$ & $203$ & $6473$ & $6473$ & $6473$ \\ \hline
\hline
CEBPB & $14500$ & $399$ & $337$ & $334$ & $8275$ & $8322$ & $8267$ \\ \hline
CTCF & $32958$ & $360$ & $492$ & $579$ & $17087$ & $17098$ & $17060$ \\ \hline
E2f4 & $4132$ & $248$ & $590$ & $517$ & $4643$ & $5146$ & $3879$ \\ \hline
Fosl1 & $5981$ & $0$ & $90$ & $90$ & $5088$ & $5088$ & $5039$ \\ \hline
Max & $8751$ & $24$ & $70$ & $81$ & $12531$ & $12495$ & $12386$ \\ \hline
MyoD & $33969$ & $717$ & $679$ & $665$ & $25416$ & $25430$ & $25344$ \\ \hline
Myog & $38292$ & $1116$ & $584$ & $835$ & $29520$ & $29334$ & $29647$ \\ \hline
NRSF & $13756$ & $639$ & $672$ & $488$ & $13183$ & $14363$ & $13440$ \\ \hline
SRF & $2370$ & $1$ & $34$ & $35$ & $2929$ & $2928$ & $2948$ \\ \hline
TCF3 & $9453$ & $185$ & $277$ & $257$ & $8528$ & $8690$ & $8775$ \\ \hline
USF1 & $8956$ & $11$ & $14$ & $12$ & $86
28$ & $8619$ & $8625$ \\ \hline
\end{tabular}
\begin{flushleft} For each TF, we show the number $N_{\text{chip}}$ of ChIP sequences retrieved, the numbers  $\Delta_{\text{inde-pairwise}}$, $\Delta_{\text{inde-mixture}}$, and  $\Delta_{\text{pairwise-mixture}}$ of different ChIP sequences used for training between either two models, and the numbers  $N_{\text{inde}}$, $N_{\text{mixture}}$, and $N_{\text{pairwise}}$ of TFBSs used to learn each model.
\end{flushleft}
\label{tab:label}
 \end{table*}

\section*{Materials and Methods}

\subsection*{Genome-wide data retrieval}

We use both ChIP-on-chip data from {\it Drosophila Melanogaster} and ChIPseq data from {\it Mus Musculus}. Data was retrieved from the litterature \cite{Zinzen:2009p760,Chen:2008p3738} and from ENCODE data accessible through the UCSC website \texttt{http://hgdownload.cse.ucsc.edu/goldenPath/ mm9/encodeDCC/wgEncodeCaltechTfbs/}, for a total of $27$ TFs. 
Among them, there are $5$ developmental Drosophilae TFs: Bap, Bin, Mef2, Tin and Twi, $11$ mammalian stem cells TFs: c-Myc, E2f1, Esrrb, Klf4, Nanog, n-Myc, Oct4, Sox2, Stat3, Tcfcp2l1, Zfx, and $11$ factors involved in mammalian myogenesis: Cebpb, E2f4, Fosl1, Max, MyoD, Myog, Nrsf, Smad1, Srf, Tcf3, Usf1. Overall, there are between $678$ and $38292$ ChIP peaks, with average size $280$bp.
DNA sequences were masked for repeats using RepeatMasker~\cite{Bao:2002uq}.

\subsection*{Background models}
It is important to discriminate  the statistics of the motifs proper from that
of the background DNA on which motifs are found. Besides particular nucleotides frequencies,  the background DNA can exhibit  significant nucleotide correlations, for instance arising  from CpG depletion in mammalian genomes (Figure \ref{backcor}). 
For each ChIPseq data, we used, as background, all sites from both strands of the sequences. This serves to learn  independent and pairwise background models
 which were used as reference models to score the corresponding TFBS models.
The position information content in all plotted PWM logos is measured with respect to the nucleotide background frequencies ({\em i.e.} the independent background model) 

\subsection*{Initial PWM refinement}

Along with the ChIPseq data for the different factors, we also retrieved corresponding PWMs from the literature \cite{Zinzen:2009p760} or from TRANSFAC database \cite{Wingender:2000uq}. These initial PWMs were refined according to the following protocol.

Given ChIPseq data (bound regions) for a given TF and an initial PWM of length $L$ ($L=12$ was taken for all computations in the present paper),
we scanned both strands of each bound region and attributed to all observed $L$-mers a score defined as the ratio between the PWM and background models probabilities. A cutoff was set such that half of the bound regions had at least one predicted TFBS with a score above the cutoff, setting a True Positive Rate (TPR) of $50\%$. This heuristic criterium overcame the problem of False Positives among the ChIPseq peaks that might have polluted the data. This defined a training set of $N$ $L$-mers with probability higher than the cutoff. Bound sites were again predicted using the same cutoff. This procedure was repeated until stabilization of the predicted sites to a fixed subset. This resulted in a refined PWM with its associated set of bound sites.

\subsection*{Independent model evaluation}

The independent model consist of a matrix of single nucleotide probabilities of size $4\times L$, where $L$ is the width of the binding site. In a first approximation, the parameters appearing in the matrix can be estimated from a set of binding sites by computing the observed frequency $f_{b,i}$ of nucleotide $b$ at position $i$. However, this frequency fluctuates around the ``true'' probability due to finite sample size, and for example unobserved nucleotides could actually have a low probability of being observed provided that the number of observations be high enough. It is usual to correct for this effect by using the Bayesian pseudo-count approach stemming from Laplace's rule of succession \cite{wasserman}. The probability to observe nucleotide $b$ at position $i$ is given by:
\begin{equation}
p_{i,b} = \frac{n_{i,b}+\alpha_b}{N+\sum_b \alpha_b}
\end{equation}
where $n_{i,b}$ is the number of observed $b$ at position $i$, $N$ is the total number of sites, and $\alpha_b$'s are the pseudo-counts, or prior probabilities to observe nucleotide $b$ at position $i$. The pseudo-counts were all set to $1$, however no significant effect was noted when changing this value, as expected from the large number of observations.

\subsection*{Kullback-Leibler divergence}
The Kullback-Leibler divergence is a measure of distance between two probability distributions $p$ and $q$ of a variable $s$, and is defined as:
\begin{equation}
\mathrm{DKL}(p\Vert q)=\sum_s p(s)\log \frac{p(s)}{q(s)}.
\end{equation}
Throughout this paper, when a DKL is calculated between a finite sample and a model distribution, $p$ corresponds to the sites frequencies in the sample, and $q$ to the model distribution. When the DKL is calculated between a PWM of a basin of attraction of a metastable state and a PWM from the mixture model, $p$ is used for the former, and $q$ for the latter.

\subsection*{Estimation of the fluctuations due to finite sampling: DKL {\em vs} self}
To estimate whether the description of the data by a model ({\em e.g.} independent or pairwise) could be improved  or was consistent with the  finite number $N$ of observed sequences, we computed the `self' DKL between the distribution of a set of $N$ sequences drawn from the model distribution and the model distribution itself.
 This procedure was repeated $100$ times. TFs for which the independent model DKL was smaller than or within two standard deviations of the self DKL were discarded for later analysis. 

\subsection*{Derivation of the pairwise interaction model}
Information theory offers a principled way to determine  the probabilities of a set of states given some  measurable constraints. It consists in maximizing
a functional known as the entropy\cite{Jaynes:1957p3771,SHANNON:1948p699} over the set of possible probability distributions given the imposed constraints. Here, we wish to determine
the probability 
$P(s)$ of a DNA sequence $s$  of length $L$,  in the set of TFBSs for a transcription factor, given the constraints that the probability distribution $P$ retrieves the one- and two-point correlations observed in a set of bound DNA sequences. We denote by $\mathcal{A}$
the alphabet of possible nucleotides, $\mathcal{A}= \{A,C,G,T\}$ and by $s_i$ the nucleotide at position i in the sequence $s$ so that $s=s_1\cdots s_L$. With these notations,  the entropy with the considered constraints  translates into the  the following functional:

\begin{widetext}

\begin{equation}
\begin{aligned}
\mathcal{L}  =  - \sum_{\{ s \}} P(s) \ln P(s) & + \lambda \left(\sum_{\{ s \}} P(s) -1\right)  +  \sum_{i=1}^L \sum_{a \in \mathcal{A}} h_i(a) \left(\sum_{\{ s \}} P(s) \delta(s_i,a) - P_i(a)\right) \\
			 & +  \sum_{i=1}^{L-1} \sum_{j>i} \sum_{a \in \mathcal{A}}  \sum_{a' \in \mathcal{A}}  J_{i,j}(a,a') \left(\sum_{\{ \sigma \}} P(a) \delta(s_i,a) \delta(s_j,s') - P_{i,j}(a,a')\right),
\end{aligned}
\label{Leq}
\end{equation}
\end{widetext}
where $ P_i(a)$ (resp. $P_{i,j}(a,a')$) is the probability  of having nucleotide $a$ at position $i$ (resp. nucleotides $a$ and $a'$ at position $i$ and $j$) in the TFBS data set.
The  function $\delta$ denotes the Kronecker $\delta-$function  defined by $\delta(a,a')=1$ if $a=a'$,and 0 otherwise.
The first term in Eq.~(\ref{Leq}) is the entropy of the probability distribution to be found and the other terms are the given constraints along with their Lagrangian multipliers. Maximization
of the functional $\mathcal{L}$ is performed in a usual way by setting the functional derivative with respect to the probability distribution $P$ to zero:
\begin{equation}
\label{eq:potts}
\frac{\delta \mathcal{L}}{\delta  P(s)}   =  0  
			  = - \ln P(s) - 1  + \lambda  
			  +  \sum_{i=1}^L h_i(s_i)
			  +  \sum_{i=1}^{L-1} \sum_{j>i} J_{i,j}(s_i,s_j).		
\end{equation}

Finally, using the constraint $\sum_{\{ s \}} P(s) = 1$, one finds the probability distribution that maximizes entropy given the constraints that it reproduces the observed one- and two-point correlations:

\begin{equation}
P[s] =   \exp[-\mathcal{H}(s)] / \mathcal{Z},
\label{proseq}
\end{equation}
where $\mathcal{H}(s)$ is the inhomogeneous Potts model Hamiltonian,
\begin{equation}
\begin{split}
\mathcal{H}[s_1...s_L] =&  - \sum_{i=1}^{L} h_i(s_i) - \sum_{i=1}^{L} \sum_{j<i} J_{i,j}(s_i,s_j),\\ &\qquad s_i \in \{A,C,G,T\}.
\label{hpotts}
\end{split}
\end{equation}
The normalization constant $\mathcal{Z}$ is the partition function,
\begin{equation}
\mathcal{Z}=\sum_{\{s\}} \exp[- \mathcal{H}(s)].
\end{equation}

\subsection*{Gauge fixing}

The probability distribution of sequences, as given by Eqs.~(\ref{proseq}, \ref{hpotts}), is invariant under shifts of the local fields $h_i(a)$  and under transformations between the interaction terms $J_{i,j}(a,a')$  and the local fields. In order to uniquely determine  $\mathcal{H}$, this arbitrariness  needs to be taken care of by adding further conditions that uniquely fix the interaction parameters, a process known  as gauge fixation \cite{Weigt:2009p3341} that we detail here.

\paragraph{Local fields.}

Since it amounts to changing  the reference energy and is cancelled by the normalization, the probability is invariant with respect to the following global shift of the $h_i(a)$ 
\begin{equation}
h_i(s_i)  \rightarrow \tilde{h}_i(s_i) = h_i(s_i) + \varepsilon_i.
\label{simh}
\end{equation}
We choose to fix this invariance by minimizing the square norm $S_{i} = \sum_{a\in \mathcal{A}} \tilde{h}_{i}(a)^2$ of local field terms with respect to the gauge degree of freedom. The corresponding gauge-fixing condition reads
\begin{equation}
\sum_{a\in \mathcal{A}} \tilde{h}_i(a) =0.
\label{condh}
\end{equation}
This condition can be imposed on any set of fields ${h}_{i}$ by using the symmetry (\ref{simh}) and redefining the fields as follows,
\begin{equation}
h_i(s_i)  \rightarrow h_i(s_i) -\frac{1}{4} \sum_{a\in \mathcal{A}} h_i(a).
\end{equation}

\paragraph{Interaction terms.}

Another invariance stems from the fact that contributions can be shifted between local fields and interaction energies. 
Namely, the following change of variables does not affect the probability:
\begin{equation}
J_{ij}(s_i,s_j) \rightarrow \tilde{J}_{ij}(s_i,s_j) = J_{ij}(s_i,s_j) + \psi_i(s_i) + \phi_j(s_j) +C_{i,j},
\end{equation}
since the local fields $\psi_i$ and $\phi_j$ can be redistributed in $h$ and the constant $C_{i,j}$ gives an energy reference for the interacting nucleotides that is cancelled by the normalization process. Again, a gauge condition is obtained by minimizing the square norm $S_{i,j} = \sum_{a,a'\in \mathcal{A}} [\tilde{J}_{ij}(a,a')]^2$ of interaction terms with respect to the gauge degrees of freedom. This yields the conditions:

\begin{equation}
\sum_{a\in \mathcal{A}} \tilde{J}_{i,j}(a,a') = \sum_{a'\in \mathcal{A}} \tilde{J}_{i,j}(a,a') = 0.
\end{equation}
These can be imposed on a set a of $J_{ij}$ parameters by redefining them as follows:
\begin{equation}
\begin{split}
J_{ij}(s_i,s_j) \rightarrow & J_{ij}(s_i,s_j) + \frac{1}{16} \sum_{a,a'\in \mathcal{A}} J_{i,j} (a,a') \\ &- \frac{1}{4} \sum_{a\in \mathcal{A}} J_{i,j}(a,s_j)  - \frac{1}{4} \sum_{a\in \mathcal{A}}  J_{i,j}(s_i,a).
\end{split}
\end{equation}

\subsection*{Determination of the pairwise interaction model from the data}

The parameters of the inhomogeneous Potts model in Eq.~(\ref{hpotts}), giving the energy of an observed sequence of length $L$, must be computed from the data.
The parameters $h$ and $J$ represent the energy contributions respectively coming from individual nucleotides and from their interactions. The PWM model is the particular case where all the interaction parameters vanish: $J_{i,j}(a,a') = 0$.

To build the model, we start from the PWM description, characterized by the set of initial $h_{i}(a)=\log p_{i,a}$ and the interaction parameters $J$'s set to zero. We add one interaction parameter $J_{i,j}(a,a')$ at a time, corresponding to the pair of nucleotides whose pairwise distribution predicted by the model differs most from data, as estimated by a binomial $p$-value. We then fit the augmented model to data, use this model to select a new set binding sites from the reads, and repeat the whole procedure.
In each of these steps, fitting is performed by a gradient descent algorithm:
\begin{eqnarray}
J&\rightarrow& J+\epsilon\left[ c_2^{\rm data}-c_2^{\rm model}\right],\\
h&\rightarrow& h+\epsilon\left[ c_1^{\rm data}-c_1^{\rm model}\right],
\end{eqnarray}
where $c_1$ and $c_2$ are matrices of size $4\times L$ and $4L \times 4L$ respectively corresponding to the single- and two-point frequencies, and superscripts denote whether the matrices are computed from the data or from the model distribution.
This algorithm converges to the set of parameters $(\{\tilde{h}_i\},\tilde{J}_{i,j})$ that match all single marginals and the pairwise marginals of interest. The number of interaction parameters that are being added is controlled by the Bayesian Information Criterion, or BIC (Figure \ref{fig:bic}).
The BIC computes the opposite log-likelihood and adds a penalty proportional to the number of parameters involved. This adverts the over-fitting of a finite dataset with an extravagant number of parameters.
The procedure is iterated until minimization of the BIC, yielding the best pairwise model with the full set of parameters $(\{h_{i}(a)\},\{J_{i,j}(a,a')\})$. As in the case of the PWM model, we score each sequence using the ratio between the TF and background pairwise models and impose a score cutoff so as to select a set of bound sites yielding $50\%$ TPR, on which a new pairwise model is learned. This process is iterated until convergence to a stable set of bound sites.

\subsection*{BIC computation}

Consider a sample $X=(X_1,\ldots,X_N)$ of $N$ TFBSs drawn from an unknown distribution function $f$ we wish to estimate. To this extent, several models $\{M_1,\ldots,M_m\}$ are proposed, each model $M_i$ having a density $g_{M_i}$ with parameter $\theta_i$ of dimension $K_i$. It is straightforward to see that, as $K_i$ increases, the fit to the observed sample as measured by the likelihood function $g_{M_i}(X|\theta_i)$ increases as well, the limiting case being when $f$ is estimated as the sample distribution. However, such an estimator is inappropriate to account for new, yet unobserved TFBSs, {\em i.e.} it is not predictive. Such a case where the number of parameters used to estimate a distribution becomes of the order of the size of the sample is known as overfitting. The BIC 
allows to overcome overfitting by penalizing high dimension parameters. Using Bayes Rule, and a uniform a priori distribution on the models, we have
\begin{equation}
P(M_i|X) \propto P(X|M_i).
\end{equation}
That is, the probability of the model given the data can be inferred from the probability that the data is generated by the model. The latter is obtained by marginalizing the joint distribution of the data and the parameters over the space of parameters $\Theta$:
\begin{equation}
P(X|M_i) = \int_\Theta P(X,\theta|M_i) d\theta = \int_\Theta g_{M_i}(X|\theta) P(\theta|M_i)d\theta.
\end{equation} 

For a unidimensional parameter $\theta$, the likelihood $g_{M_i}(X|\theta)$ is maximized at some particular $\hat{\theta_i}$ with an uncertainty (or width) proportional to $1/\sqrt{N}$ in the limit of large $N$. Assuming a broad prior, then for large $N$ the integral is dominated by the likelihood which is concentrated around its maximum. One can then approximate the integral by the area of the region of height the maximum likelihood and of width $1/\sqrt{N}$, that is $g_{M_i}(X, \hat{\theta_i})/\sqrt{N}$. This result can be retrieved analytically using the method of steepest descent. For a number $K_i$ of parameters, one gets a total volume $g_{M_i}(X, \hat{\theta_i})/N^{K_i/2}$ \cite{bishop}. Taking the logarithm yields the BIC condition:
\begin{equation} 
BIC_i = - 2 \log( P(X | M_i) ) \simeq - 2 \log(g_{M_i}(X, \hat{\theta_i})) + K_i \log(N).
\label{bic}
\end{equation}
In the present case, the sample $X$ is the set of observed TFBSs and the model $M_i$ determines the probability $P_{M_i}(s)$ of belonging to $X$, 
\begin{equation}
 \log(g_{M_i}(X, \hat{\theta_i}))= \sum_{s \in X} \log[P_{M_i(\hat{\theta_i})}((s)].
\end{equation}

The interpretation of Eq.~(\ref{bic}) is 
clear: adding new parameters improves the fit, but also adds new sources of uncertainty about these parameters due to the finite size of the data. This uncertainty disappears as $N \to \infty$, since the log-likelihood scales with $N$ while the correction scales with $\log(N)$.

Finally, Eq.~(\ref{bic}) is a functional over models, the chosen model $M_{BIC}$ is the one that minimizes it, 

\begin{equation} 
M_{BIC}=
\underset{M_i}{\mathrm{argmin}}\ BIC_i.
\end{equation}

\subsection*{PWM mixture model}
 We investigated an approach based on a mixture of PWMs. For that purpose, we used a comparable setup as for the pairwise model. However, instead of adding correlations to a given PWM, new PWMs were added to a mixture model. More precisely, a mixture of $K$ PWMs, with $1\leq K\leq10$, was generated by using a K-means algorithm with a Hamming distance metrics on the initial set of bound sites. This resulted in $K$ clusters, each comprising $n_k$ sites among the initial $N$ sites. A PWM was generated on each of these clusters, with probability distribution $\mathcal{P}_k$. The mixture model of order $K$ was then defined as \cite{bishop}:

\begin{equation}
\mathcal{P}[s] =  \sum_{k=1}^{K} p_k \mathcal{P}_k[s],
\end{equation}
where $p_k=n_k/N$ is the cluster weight. Because a PWM has $3\times L$ degrees of freedom ($L$ of them being constrained by the summation of nucleotide probabilities to one) and there are $K-1$ free weight parameters, the number of parameters corresponding to a mixture of order $K$ is $3LK + (K - 1)$. As previously, the model showing minimal BIC score was used for sites detection, a new set of PWMs  and weights $p_k$ was generated by clustering the set of detected sites  and the procedure was iterated until convergence to a stable set of sites.

\subsection*{Metastable minima of the pairwise interaction model and their basins of attractions}
We defined the basins of attraction of a pairwise interaction model energy landscape, in the following fashion. Let $s$ be a site with energy $\mathcal{H}(s)$. We looked for the nucleotides that could be changed to minimize $\mathcal{H}(s)$. If such nucleotides existed, one of them was chosen at random, and its value was updated. One local minimum of the energy landscape, or metastable state, was reached when no such nucleotide could be found. The basin of attraction of a metastable state was then defined as the ensemble of sites that fell to this metastable state when their energy was minimized following the above procedure. We computed metastable states and their basins of attraction for the final set of bound sites obtained with the best pairwise model. A PWM was learned on each basin of attraction, leading to a set of representative PWMs, with different weights representing different proportions of bound sites in their basins.

\subsection*{Computation of the Direct Information}

We wanted to build a quantity based solely on direct interactions $J_{i,j}$ between nucleotides, discarding indirect interactions.
To this end, we used the interaction parameters obtained from the pairwise model to build the direct dinucleotide probability function:
\begin{equation}
 P_{i,j}^d(a,a') =  e^{ \tilde{h}_i(a) + \tilde{h}_j(a') + J_{i,j}(a,a')} /  \mathcal{Z}_{i,j},
 \end{equation}
where $$ \mathcal{Z}_{i,j}=\sum_{a,a'}   e^{ \tilde{h}_i(a) + \tilde{h}_j(a') + J_{i,j}(a,a')}.$$

The $8$ effective fields $\tilde{h}_i$ and $\tilde{h}_{j}$ were fully determined by the constraints that the direct probability function matches the observed one-point frequencies:
\begin{equation}
 \begin{aligned}
 \sum_{a'} P_{i,j}^d(a,a') &= P_i(a), \qquad a' \in \{A,C,G,T\},\\
 \sum_{a} P_{i,j}^d(a,a') &= P_j(a'), \qquad a \in \{A,C,G,T\}.
 \end{aligned}
 \end{equation}
The normalization of the probabilities $\sum_{a} P_i(a) = 1$,  served to reduce this system to $6$ equations. The fields $\tilde{h}_i(a)$, which are determined up to a constant, were fixed by the gauge condition that they vanished for the nucleotide $A$, $\tilde{h}(A)=0$. The system was solved using the Levenberg-Marquadt algorithm with $\lambda=0.005$.

The Direct Information \cite{Morcos:2011p12728}  was then defined as:
\begin{equation}
DI_{i,j} = \sum_{a,a'} P_{i,j}^d(a,a') \log_2\left(\frac{ P_{i,j}^d(a,a')}{P_i(a)P_j(a')}\right).
 \end{equation}

As there is no upper bound for this direct information, we 
built a normalized version of the direct information:
\begin{equation}
NDI_{i,j} = \frac{DI_{i,j}}{\sqrt{S_iS_j}},
 \end{equation}
where $S_i$ denotes the entropy at position $i$. Note that $S_i= DI_{i,i}$ so that $NDI_{i,i}=1$ for this maximally correlated case. On the contrary, independent nucleotides give $NDI_{i,j}=DI_{i,j}=0$. 

\subsection*{Participation Ratio}

 For each TF, an interaction weight was defined for each pair of nucleotides as 
\begin{equation}
w_{i,j} = NDI_{i,j} / \sum_{i \neq j}NDI_{i,j}.
\end{equation}
Self-interactions have no meaning here and were attributed $w_{i,i}=0$. Let us note $N= L(L-1)$ the number of possible interactions. Using our weight, one writes the  Participation Ratio as:
\begin{equation}
R = \frac{1}{ N\sum_{i \neq j}w_{i,j}^2}.
\end{equation}

The interpretation is simple: if all weights are equal, $w_{i,j}=1/N$ and $R=1$, that is all possible interactions are represented. Conversely, if only one interaction accounts in the distribution budget, then $R=1/N$, meaning that only one of all possible interactions is represented. 

\subsection*{Distance between interactions}

The previously defined interaction weights were averaged over all possible pairs of nucleotides at a given distance $d$ of one another, yielding the distance distribution:

\begin{equation}
P(|i-j|=d) = \mathcal{Z}^{-1}\frac{1}{N-d} \sum\limits_{|i-j|=d} w_{i,j},
\end{equation}
where
\begin{equation}
\mathcal{Z}=\sum\limits_{d=1}^{N-1}\frac{1}{N-d} \sum\limits_{|i-j|=d} w_{i,j}
\end{equation}
is a normalization factor. Note that we introduced a correction $1/(N-d)$ to account for finite-size effects, namely the fact that randomly distributed interactions will lead to an overrepresentation of nearest neighbours interactions just because these are more numerous.

\subsection*{Interaction matrix and Hopfield patterns}

In the Hamiltonian shown in \eqref{pottsh}, only $16 L(L-1)/2$ terms appear in the interaction budget: indeed, we forbid self-interations (already accounted for by the local field $h$) and do not count the interactions twice. However, we can straightforwardly extend the interaction matrix to a full symmetric matrix $\hat{J}_{(i,a),(j,b)}$ of size $(4L)^2$, with $4L$-valued indices $(i,a), i\in \{1,\cdots,L\}, a \in \mathcal{A}$. The matrix $\hat{J}$ is such that  for $i>j, \hat{J}_{(i,a)(j,b)}= J_{i,j}(a,b)$ with furthermore
$\hat{J}_{(i,a),(i,b)}=0$ and $\hat{J}_{(i,a),(j,b)}=\hat{J}_{(j,b),(i,a)}$.
The energy of a sequence $s$ can then be written with these notations
\begin{equation}
\sum_{1\leq i < j \leq L} J_{i,j}(s_i,s_j) = \frac{1}{2}\sum_{i=1}^{L} \sum_{j=1}^{L} \hat{J}_{(i,s_i),(j,s_j)}= v(s)^{\dagger} \hat{J} v(s),
\label{senmat}
\end{equation}
where in the last equality  the $^{\dagger}$ sign denotes  vector transposition and we have introduced the $4L$ vector $v(s)$ associated to sequence $s$, $v(s)_{i,a}=1$ if $a=s_i$ and  $v(s)_{i,a}=0$ otherwise.

Since the matrix $\hat{J}$ is symmetric, it can be diagonalized in an orthonormal basis of eigenvectors $\xi^k,\ k=1,\cdots, L$ 
with real eigenvalues
$\lambda_k$,
\eqn{
\hat{J}=\sum_k \lambda_k \xi^k\ \xi^{k \dagger}.
}
 Denoting by $\xi^k_{(i,a)}$ the coordinates of the k-th eigenvector 
then, one can rewrite Eq.~(\ref{senmat}) as
\begin{equation}
\sum_{1\leq i < j \leq L} J_{i,j}(s_i,s_j) = \frac{1}{2}\sum_{k=1}^{4L} \lambda_k \left(\sum_{i=1}^L \xi_{(i,s_i)}^k\right)^2.
\end{equation}
Finally, the full
Hamiltonian is given by:

\eqn{
\mathcal{H}=-\sum_i h_i(s_i) - \frac{1}{2}\sum_{k=1}^{4L} \lambda_k \left(\sum_{i=1}^L \xi_{(i,s_i)}^k\right)^2.
}

\section*{Acknowledgments}
We wish to thank PY Bourguignon and I Grosse for stimulating discussions at a preliminary stage of this work.

\bibliographystyle{apsrev}
\bibliography{marc,thierry,vh}

\appendix

 \clearpage
 \newpage
 
\onecolumngrid

\noindent \textbf{\Large Supporting Figures}
\vskip .5cm
\makeatletter
\renewcommand{\thefigure}{
S\@arabic\c@figure}
\makeatother
\setcounter{figure}{0}

\begin{figure}[h]
\begin{center}
\includegraphics[width=0.8\linewidth]{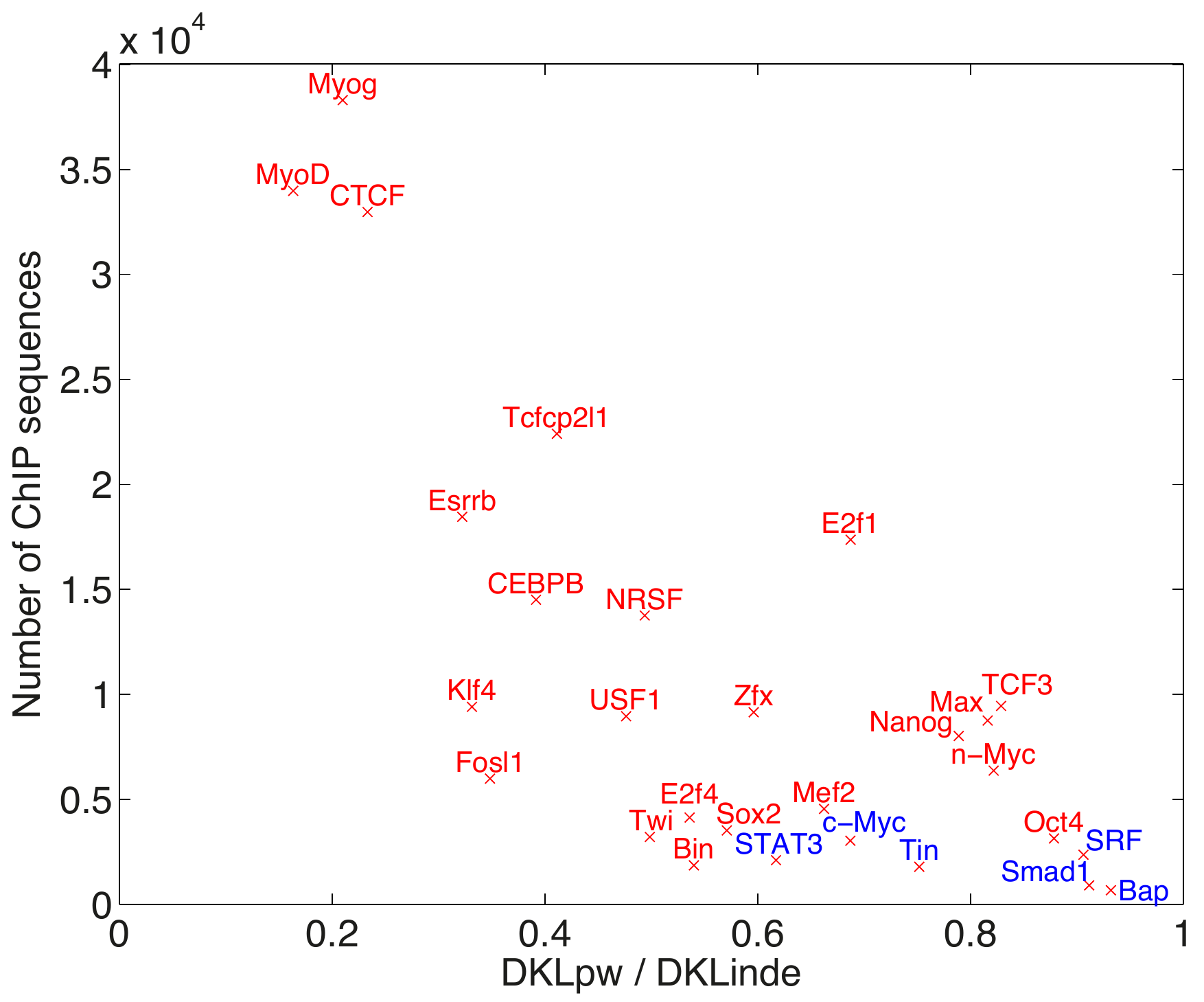}
\end{center}
\caption{{\bf Dependence of the fit on the number of ChIP sequences}. For each TF, the number of available ChIP sequences is plotted {\em vs.}   the  improvement in the description of its TFBS statistics,  provided by the he pairwise model as compared to the PWM independent model. The latter is  quantified by the ratio of DKL between the respective model probability distributions and the experimental ones provided by the ChIP data, $\text{DKL}_{\text{pw}} /  \text{DKL}_{\text{inde}}$. The improvement afforded by the pairwise model is clearly seen to be correlated to the number of ChIP sequences available.  
}
\label{corrseqnumbimprov}
\end{figure}

\begin{figure*}[!h]
\begin{center}
\includegraphics[width=\linewidth]{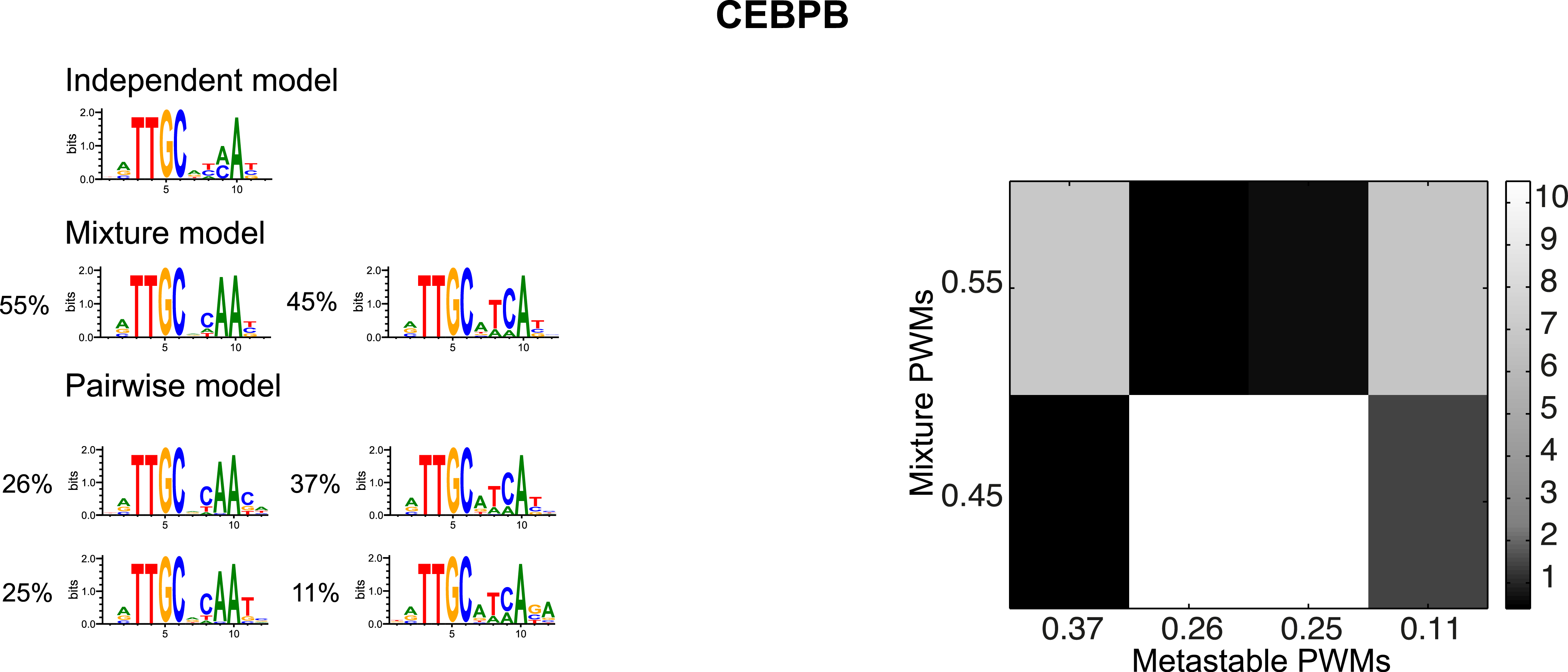}
\end{center}
\label{figS2-cebpb}
\end{figure*}
\begin{figure*}[!ht]
\begin{center}
\includegraphics[width=\linewidth]{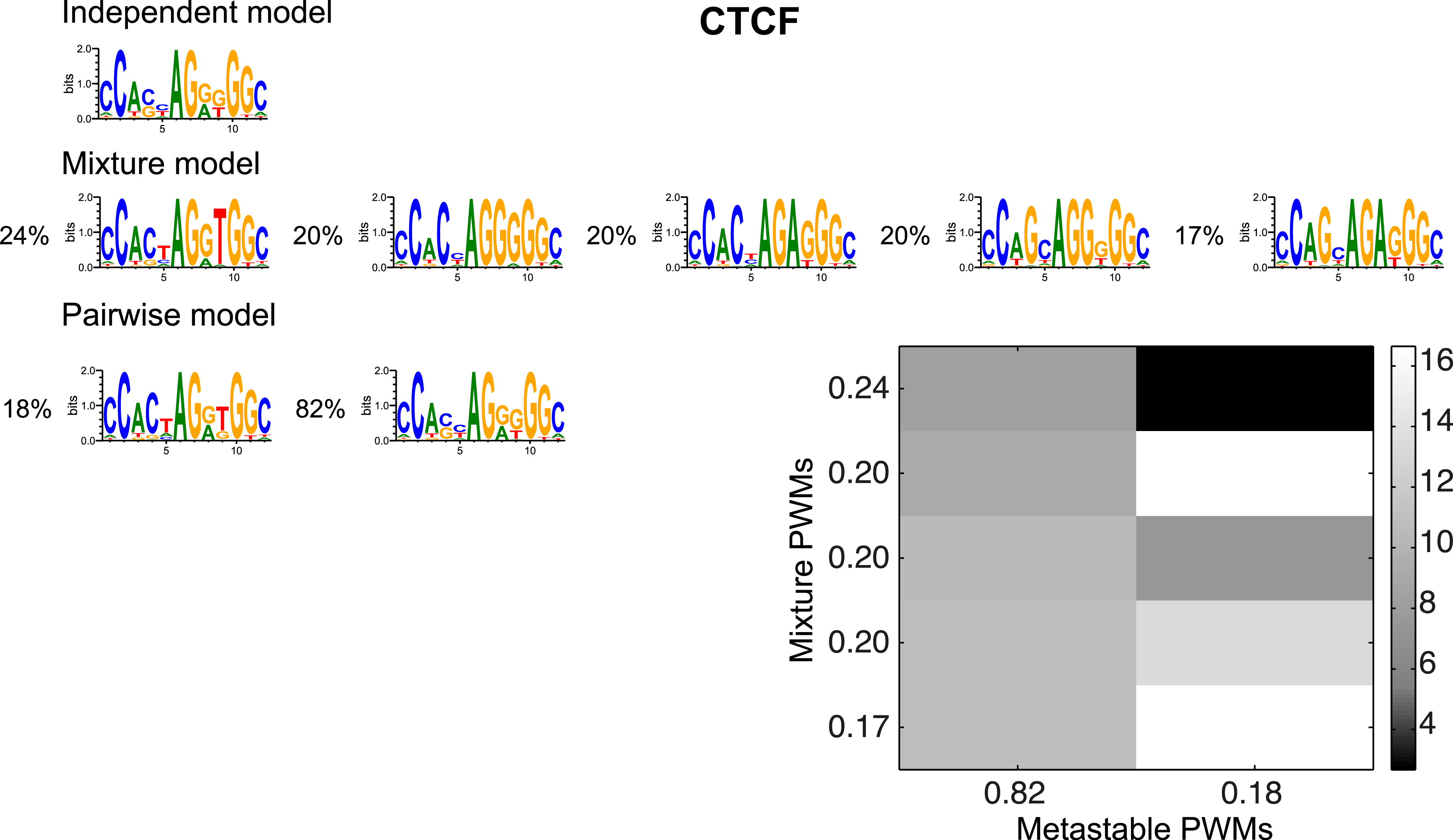}
\end{center}
\label{figS2-ctcf}
\end{figure*}

\begin{figure*}[!ht]
\begin{center}
\includegraphics[width=\linewidth]{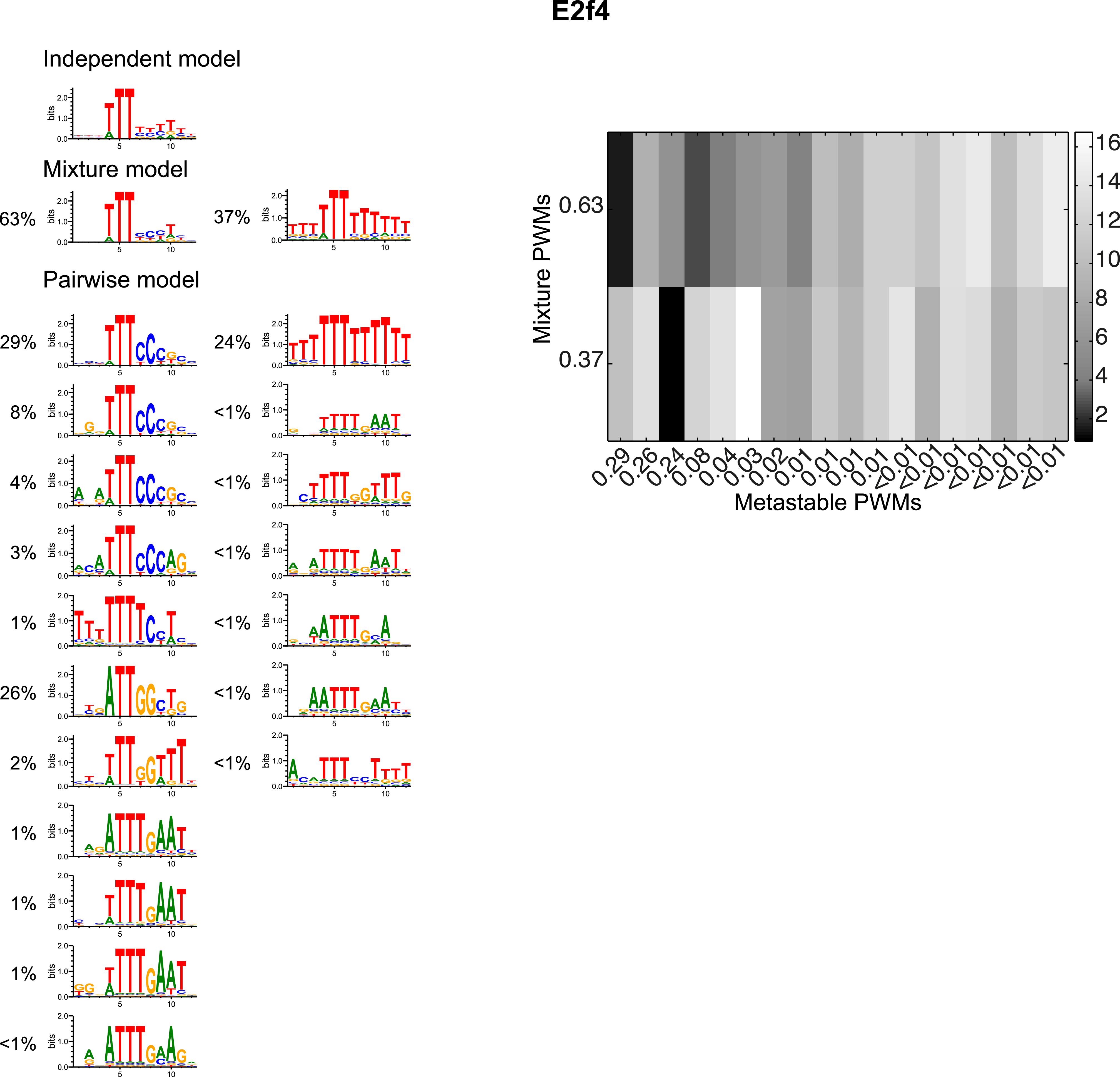}
\end{center}
\label{figS2-e2f4}
\end{figure*}

\begin{figure*}[!ht]
\begin{center}
\includegraphics[width=\linewidth]{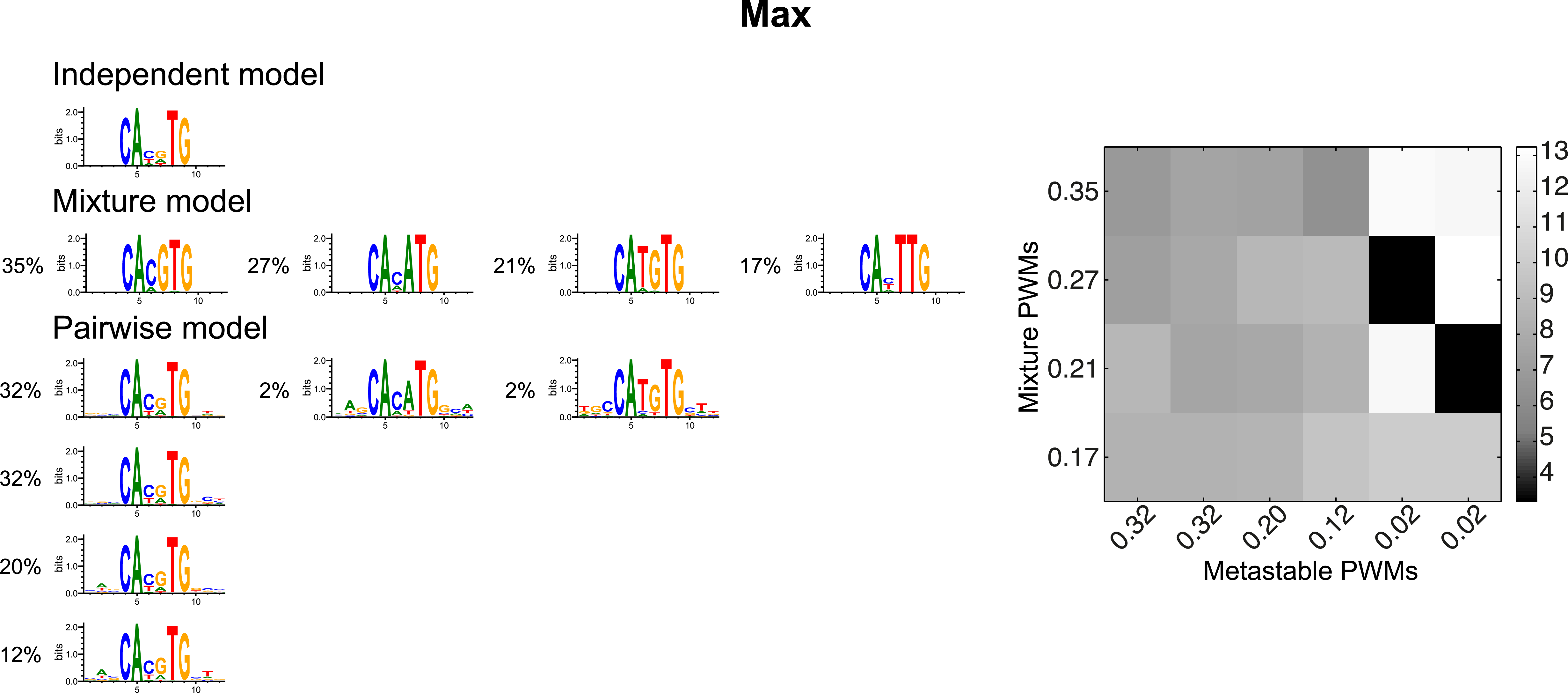}
\end{center}
\label{figS2-max}
\end{figure*}

\begin{figure*}[!ht]
\begin{center}
\includegraphics[width=\linewidth]{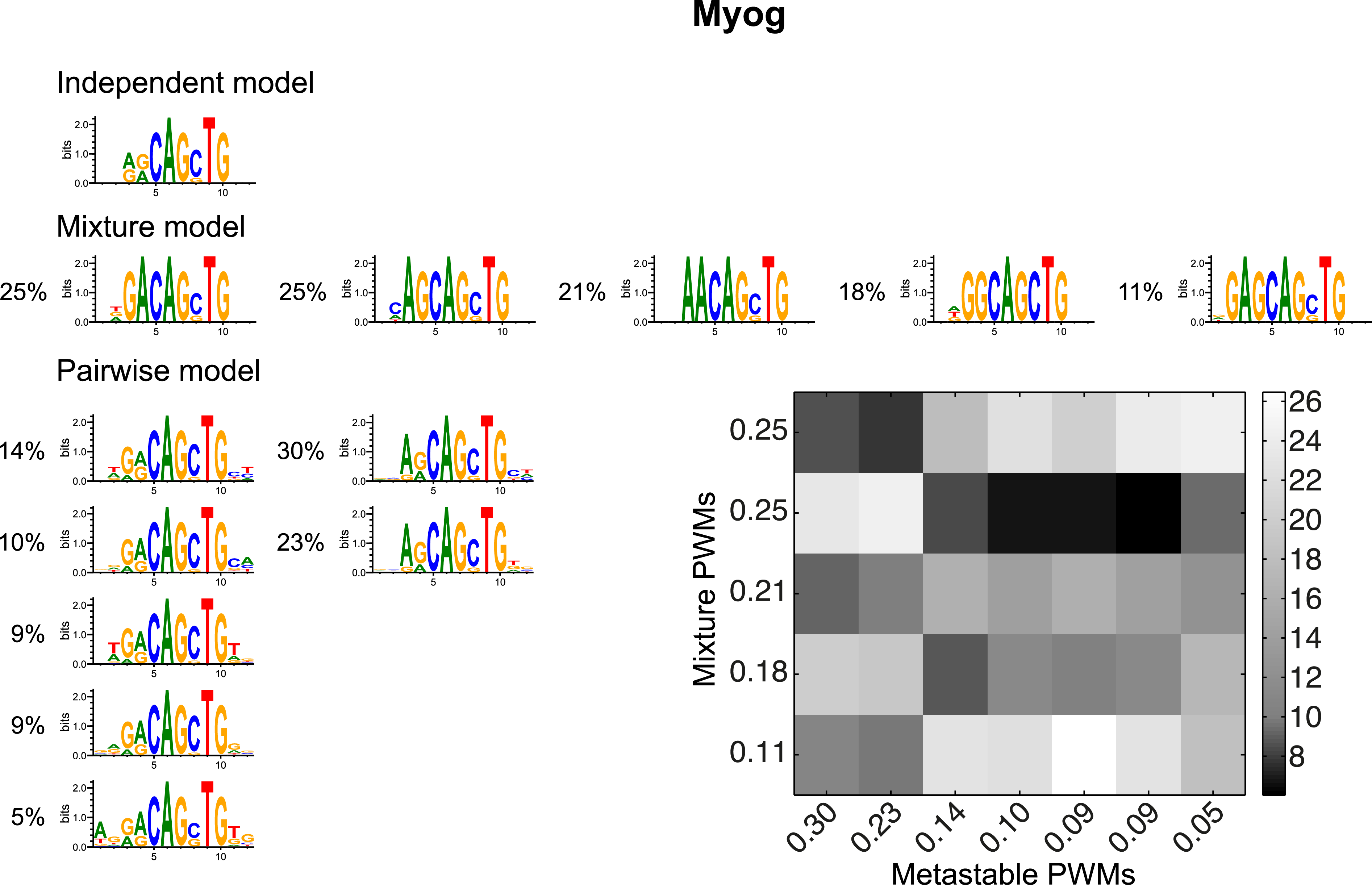}
\end{center}
\label{figS2-myog}
\end{figure*}

\begin{figure*}[!ht]
\begin{center}
\includegraphics[width=0.8\linewidth]{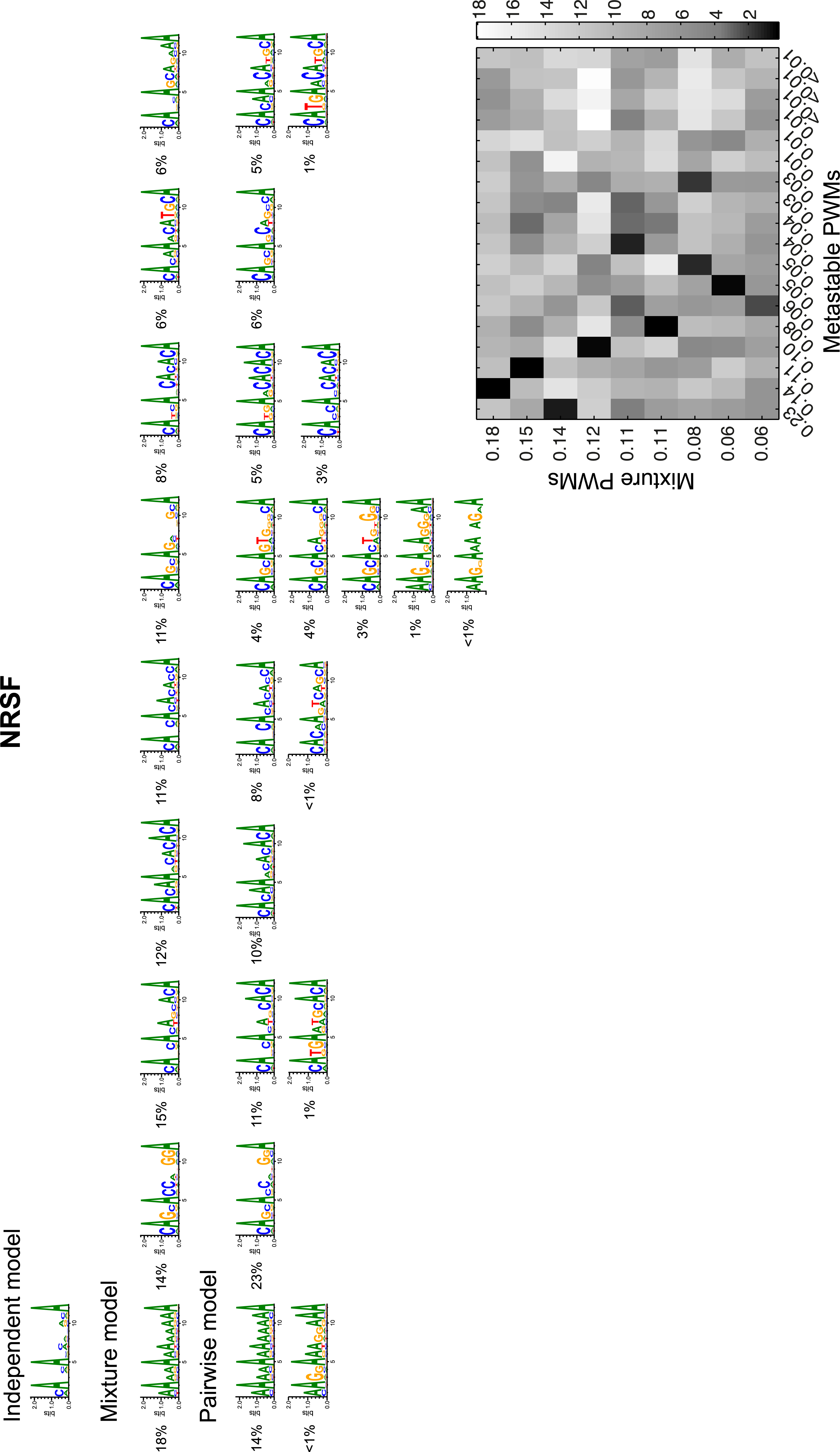}
\end{center}
\label{figS2-nrsf}
\end{figure*}

\begin{figure*}[!ht]
\begin{center}
\includegraphics[width=\linewidth]{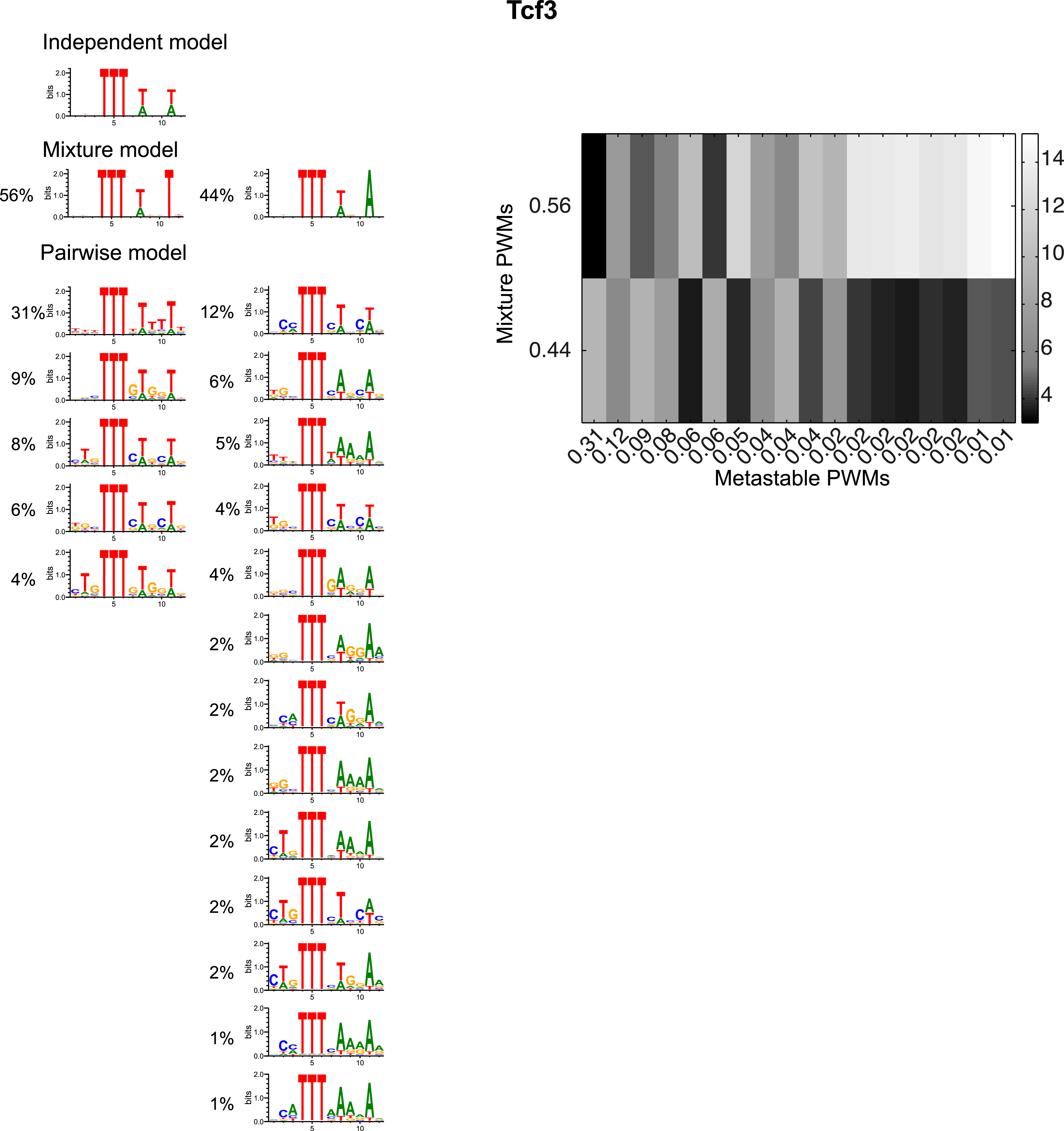}
\end{center}
\label{figS2-tcf3}
\end{figure*}

\begin{figure*}[ht]
\begin{center}
\includegraphics[width=\linewidth]{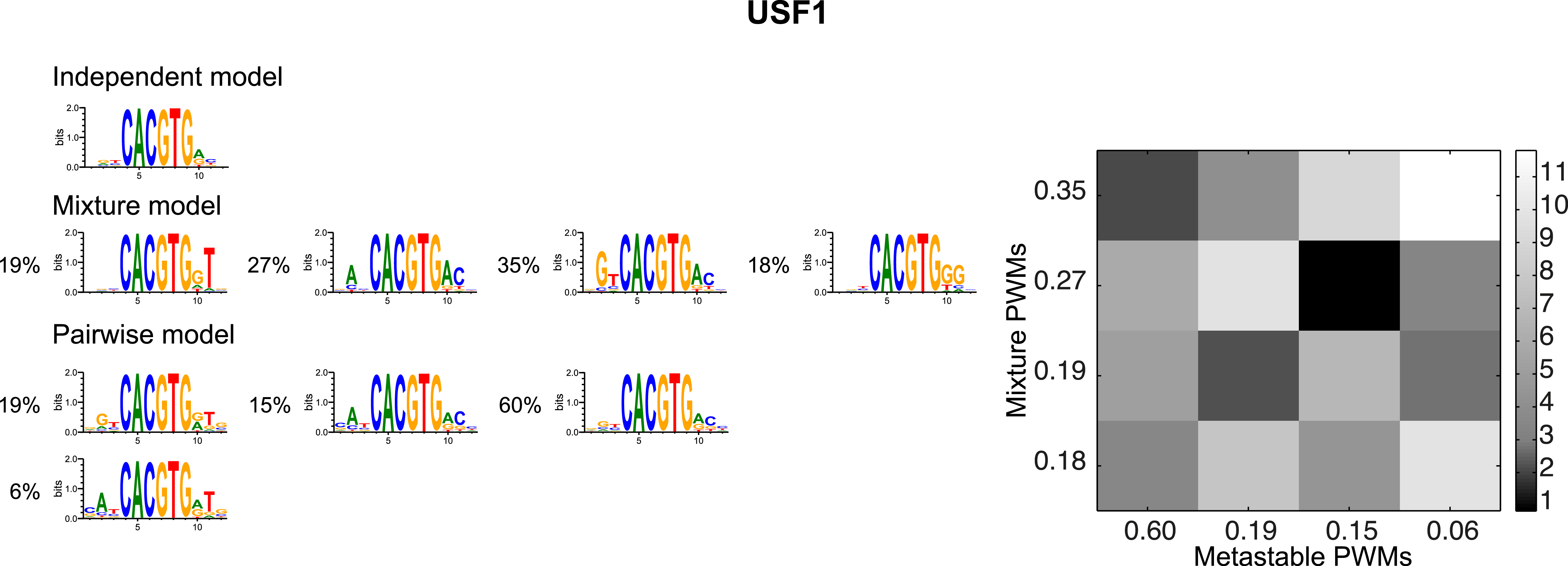}
\end{center}
\caption{Same as Figure \ref{fig:metastable} of the main text for all considered factors described by a mixture model with two or more PWMs.}
\label{figS2-usf1}
\end{figure*}

\begin{figure}[!ht]
\begin{center}
\includegraphics{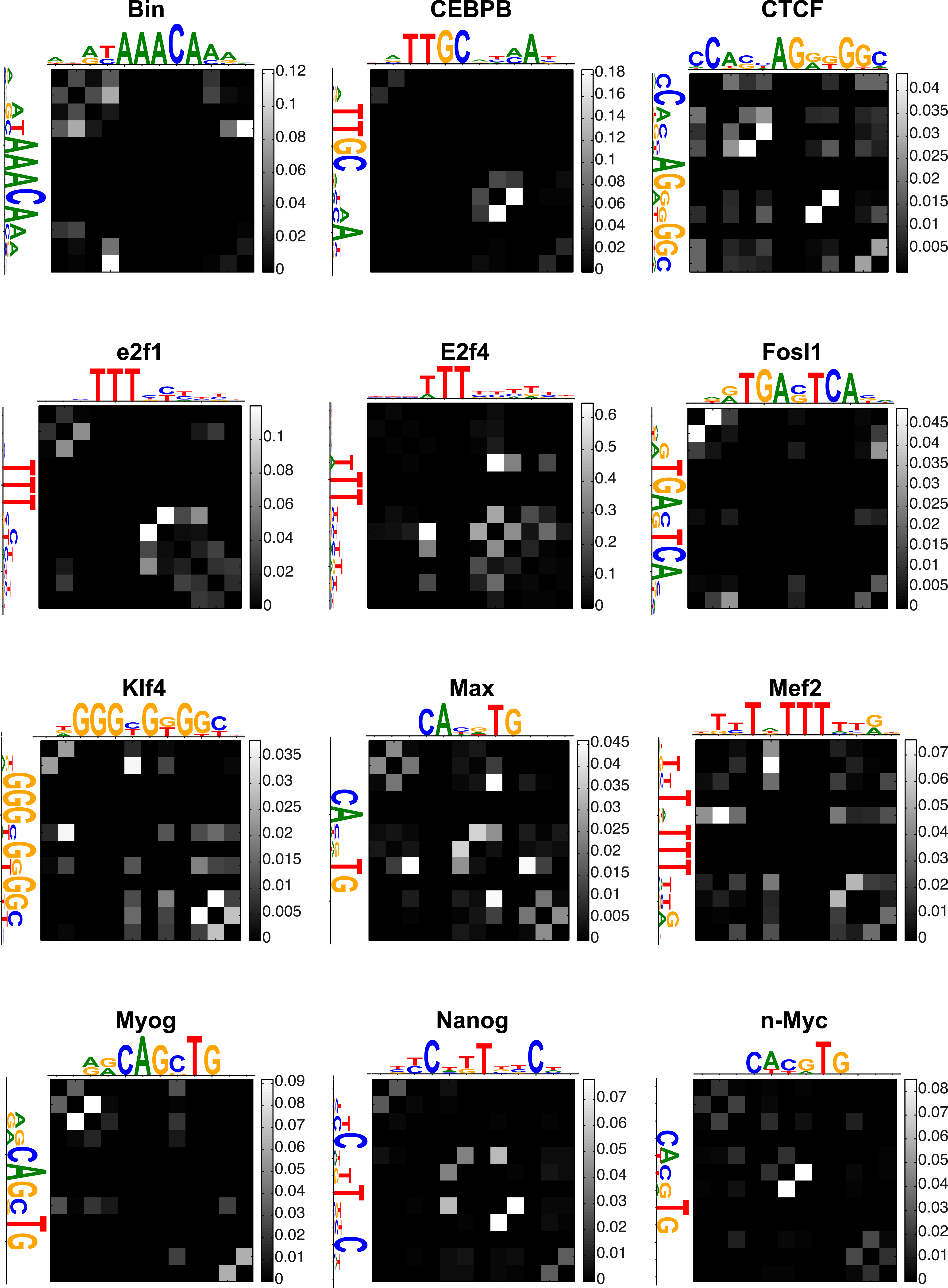}
\end{center}
\label{fig:dirinfoSI1}
\end{figure}

\begin{figure*}[ht]
\begin{center}
\includegraphics{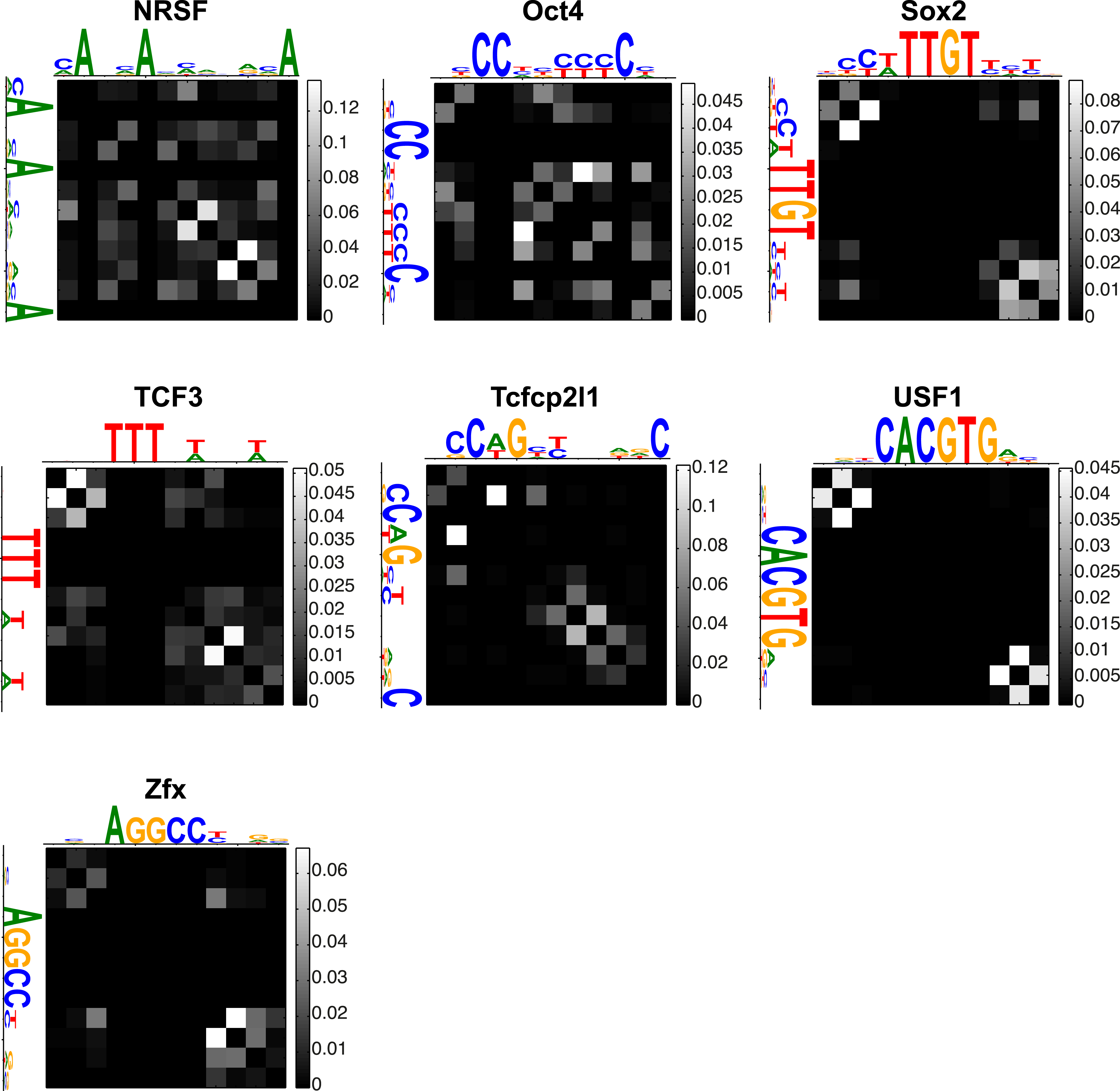}
\end{center}
\caption{Same as Figure \ref{fig:dirinfo} of the main text for the other considered factors.}
\label{fig:dirinfoSI2}
\end{figure*}

\begin{figure*}[ht]
\begin{center}
\includegraphics[width=.8\linewidth]{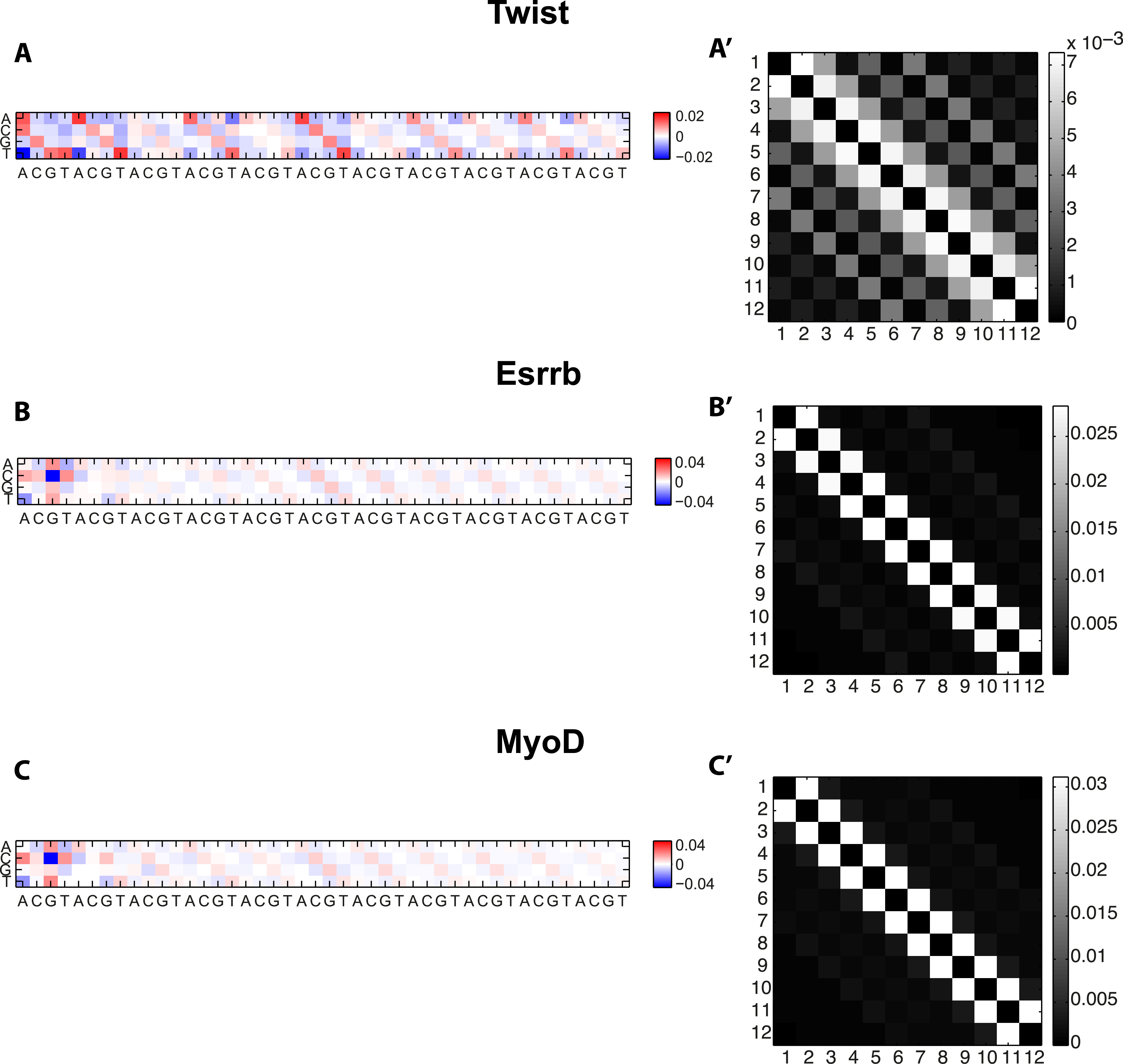}
\end{center}
\caption{{\bf Background correlations} (A,B,C) Heat maps showing the correlations between nucleotides in the ChIP data of the $3$ factors from the main text. Because of translation invariance, we only show the correlations between a nucleotide (rows) and the next nearest (first four columns) to farthest (last four columns) nucleotides, using the binding site length of $L=12$. We see in the Drosophila data the appreciable presence of repeated sequences (of type AA, TT, CC, and GG). In the mammalian data sets, we observe the known CpG depletion. (A',B',C')  Heat maps showing the values of the Normalized Direct Information between pairs of nucleotides.
}
\label{backcor}
\end{figure*}
 
\begin{figure*}[ht]
\begin{center}
\includegraphics[width=.8\linewidth]{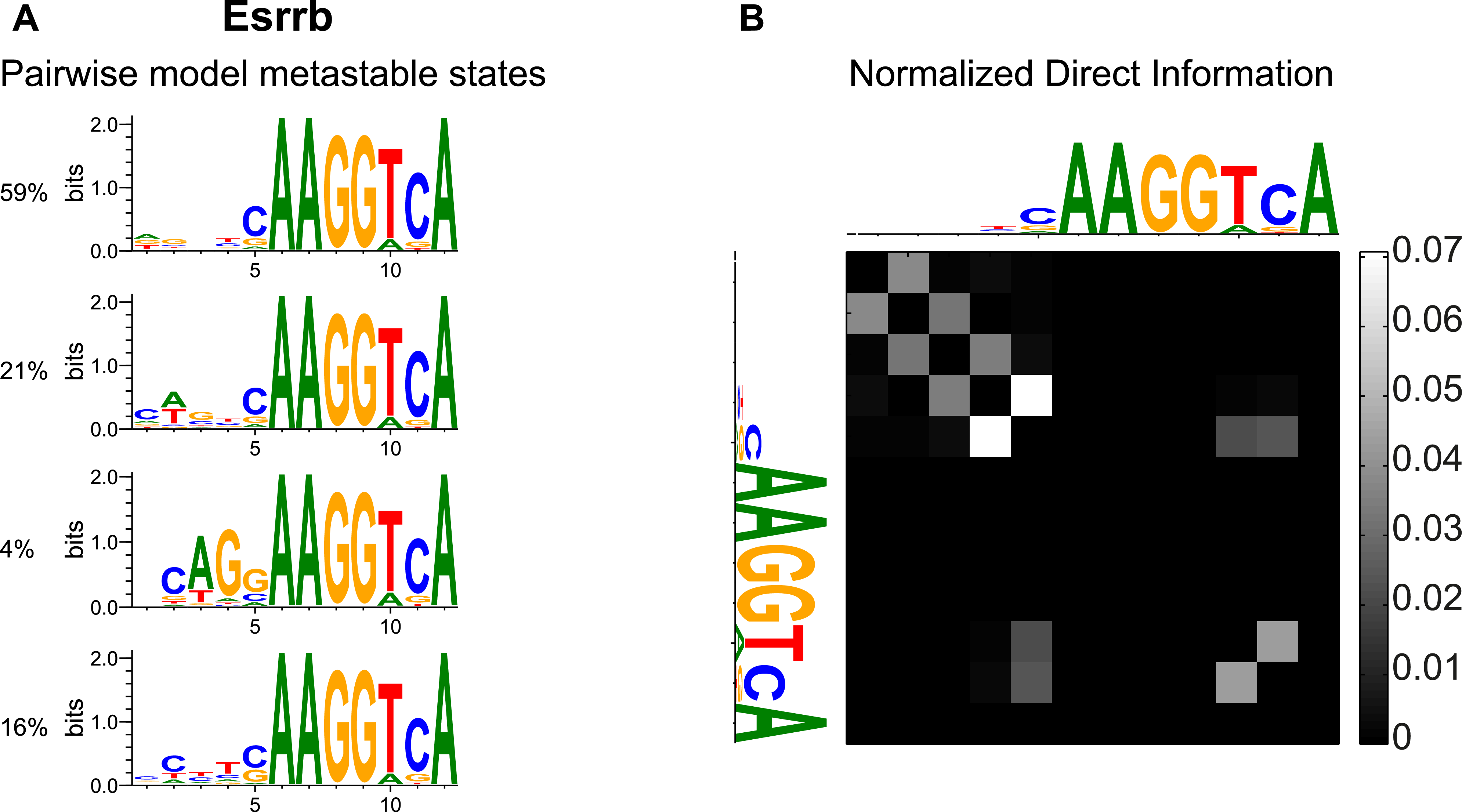}
\end{center}
\caption{{\bf Variable spacer length} We learned a pairwise model for Esrrb including the $4$ flanking nucleotides on the left of the main motif. (A) The metastable states of this model show a feature not captured in the main text where binding sites are defined symmetrically around the center of mass of the information content: namely a `CAG' trinucleotide with variable spacer length from the main motif. This feature is apparent in the first $3$ logos shown here. (B) The contribution of this trinucleotidic interaction to the Direct Information is captured through strong direct links between the $4$ flanking nucleotides, showing that the pairwise model is implicitly able to capture higher order correlations. Logos from the PWM model are surrounding the heatmap for clarity.
}
\label{figspacer}
\end{figure*}

\end{document}